\documentclass[journal]{IEEEtran}
\IEEEoverridecommandlockouts
\ifCLASSINFOpdf
\else
   \usepackage{graphicx}
   \graphicspath{{../eps/}}
   \DeclareGraphicsExtensions{.eps}
\fi
\usepackage{amsthm,amsmath,amssymb}
\usepackage{mathrsfs}
\usepackage{amsfonts}
\usepackage{graphicx, amssymb}
\usepackage{amsthm}
\usepackage{amsmath}
\usepackage{epsfig}
\usepackage{graphicx}
\usepackage{graphics}
\usepackage{multirow}
\usepackage{threeparttable}
\usepackage{amsthm}
\usepackage{float}
\usepackage{color}
\usepackage{algorithm}
\usepackage{algpseudocode}
\usepackage{bm}
\usepackage{algorithmicx}
\usepackage{algpseudocode}
\usepackage{setspace}
\usepackage{authblk}
\usepackage{cite}
\usepackage{pifont}
\usepackage{xcolor}
\usepackage{epstopdf}
\usepackage{caption}
\usepackage{graphicx}
\usepackage{float}
\usepackage{subcaption}
\usepackage{cleveref}
\theoremstyle{definition}

\newcommand{\RNum}[1]{\uppercase\expandafter{\romannumeral #1\relax}}
\usepackage{wrapfig}
\usepackage{psfrag}
\usepackage{graphicx}
\usepackage{threeparttable}
\usepackage{cases}
\usepackage{subeqnarray}
\usepackage{color}
\usepackage{underscore}

\usepackage{textcomp}
\usepackage{xcolor}
\def\BibTeX{{\rm B\kern-.05em{\sc i\kern-.025em b}\kern-.08em
    T\kern-.1667em\lower.7ex\hbox{E}\kern-.125emX}}

\newcounter{TempEqCnt} % 创建临时变量TempEqCnt
\begin{document}
%\title{How much training is needed in Integrated Super-resolution Sensing and Communication-based Symbiotic Radio Systems?}
\title{Integrated Super-resolution Sensing and Symbiotic Communication with 3D Sparse MIMO for Low-Altitude UAV Swarm}

\author{{Jingran~Xu{*}, Hongqi~Min{*}, Yong~Zeng{*$\dagger $}}\\
{{*}National~Mobile~Communications~Research~Laboratory,~Southeast~University,~Nanjing~210096,~China}\\
{{$\dagger $}Purple~Mountain~Laboratories,~Nanjing~211111,~China}\\
Email:~{\{jingranxu,~minhq,~yong_zeng\}}@seu.edu.cn}
\author{
Jingran Xu, Hongqi Min, Yong Zeng, \IEEEmembership{Fellow,~IEEE}

        % <-this % stops a space
\thanks{Part of this work will be presented at the 2025 IEEE 101st Vehicular Technology Conference (VTC), Oslo, Norway, 17-20 Jun. 2025 \cite{vtc}.}
\thanks{The authors are with the National Mobile Communications Research
Laboratory, Southeast University, Nanjing 210096, China. Yong Zeng is also
with the Purple Mountain Laboratories, Nanjing 211111, China (e-mail:
{\{jingranxu, minhq, yong_zeng\}}@seu.edu.cn). (\emph{Corresponding author: Yong Zeng.})}
}
%        % <-this % stops a space
%\thanks{%This work was supported by the National Key R\&D Program of China with
%%grant number 2019YFB1803400, by the Natural Science Foundation of China under Grant 62071114, and also by the ``Program for Innovative Talents and Entrepreneur in Jiangsu" under grant number 1104000402.
%
%Jingran Xu, Zhuoyin Dai, Yong Zeng, and Shi Jin are with the National Mobile Communications Research
%Laboratory, and Frontiers Science Center for Mobile Information Communication
%and Security, Southeast University, Nanjing 210096, China. Y. Zeng is also
%with the Purple Mountain Laboratories, Nanjing 211111, China (e-mail:
%{jingran_xu,~zhuoyin_dai,~yong_zeng,~jinshi}@seu.edu.cn). (\emph{Corresponding author: Yong Zeng.})
%
%Tao Jiang is with the Research Center of 6G Mobile Communications,
%School of Cyber Science and Engineering, and Wuhan National Laboratory for Optoelectronics, Huazhong
%University of Science and Technology, Wuhan 430074, China (e-mail: tao.jiang@ieee.org).}}
\maketitle

% As a general rule, do not put math, special symbols or citations
% in the abstract or keywords.
% \vspace{-0.5cm}

% \vspace{-0.3cm}

%\begin{itemize}
%	\item Different from prior works in angle-aided channel estimation \cite{DFT,music1,music2,music3,ESPRIT}, which are based on dedicated pilot signals.
% In our proposed method, rather then consuming a priori known pilots, both the existing pilot and data symbols can be utilized to estimate the angles of the multi-path channel components based on subspace-based super-resolution algorithms.
%	\item Based on the fact that the path AoA $\bm \Theta$ remains unchanged during several coherence time, the estimation result in the former coherence blocks is still valid for the following coherence time.
%	\item With angles obtained during each path-invariant block, the beamforming gain
%can be enjoyed when pilots are sent to estimate the channel path gains $\bm \alpha$, which can further improve the accuracy of channel estimation.
%\end{itemize}
\begin{abstract}
Low-altitude unmanned aerial vehicle (UAV) \mbox{swarms} are expected to play important role for future intelligent aerial systems due to their great potential to cooperatively accomplish complicated
missions effectively.
However, there are important challenges to be addressed to enable their efficient operation: the large-scale nature of swarms usually leads to excessive spectrum consumption,
and ultra-low-cost requirements for individual UAVs renders it necessary to develop more cost-effective communication modules.
In addition, the densely located swarm UAVs require high resolution for localization and sensing.
In order to address the above challenges and simultaneously achieve spectrum- and energy-efficient communication and accurate sensing,
we investigate low-altitude UAV swarm with integrated super-resolution sensing and symbiotic communication technology.
Specifically, one leading UAV may act as a primary transmitter (PT) to transmit communication signals to the base station (BS),
and the remaining nearby UAVs in the swarm act as passive backscatter devices (BDs),
which can modulate their information by efficiently backscattering the radio frequency (RF) signals from the PT without consuming extra spectrum or power.
In addition, to achieve efficient three-dimensional (3D) super-resolution sensing for the densely located UAV swarm, 3D
sparse multiple-input multiple-output (MIMO) technology and super-resolution signal processing algorithms are further exploited,
where both L-shaped nested array (LNA) and planar nested arrays (PNA) are considered at the BS.
To evaluate the communication and sensing performance for the UAV-symbiotic radio (SR) system,
the achievable rates of UAV swarm are derived and the beam patterns of sparse LNA, PNA and the benchmarking compact uniform planar array (UPA) are compared.
Furthermore, efficient channel estimation methods assisted by super-resolution sensing are proposed.
Simulation results are provided to demonstrate that 3D sparse MIMO with both LNA and PNA
can provide significantly better sensing and communication performance than conventional compact MIMO for UAV-SR systems.
\end{abstract}

\IEEEpeerreviewmaketitle

\section{Introduction}
Low-altitude unmanned aerial vehicles (UAVs) have been widely used in the fields
of military, civilian and commercial domains due to their flexibility and cost-effectiveness \cite{UAV1,UAV2,UAV3}.
However, due to the limitations of weak task
execution ability and low efficiency, it is difficult for one single UAV to complete complex missions.
To overcome this challenge, UAV swarms are expected to play important roles for future intelligent systems, where multiple UAVs work as a team to accomplish complex tasks cooperatively.
UAV swarm offers enhanced intelligence, improved coordination, increased flexibility, survivability, and
reconfigurability \cite{UAVswarm1,UAVswarm2,UAVswarm3}, which has a variety of applications such as rescue operations, heavy cargo delivery, pollution monitoring, smart factory,
coordinated light shows and border monitoring \cite{UAVswarm3}.

However, there are multiple challenges to be addressed before large-scale deployment of UAV swarms in communication systems.
First, the massive number of UAVs in a swarm leads to surging spectrum demands,
where traditional spectrum utilization struggles to adapt to dynamic high-density scenarios, resulting in resource competition and interference.
Furthermore, to achieve scalable swarm applications, the hardware cost of a single UAV needs to be minimized, but traditional communication architectures that rely on costly radio frequency (RF) modules and independent energy supplies hinder cost efficiency.
In addition, densely located swarm UAVs exacerbate the challenge of active localization or passive sensing, which requires super-high spatial resolutions.
To address these challenges, symbiotic radio (SR) offers a novel solution by enabling spectrum and energy sharing between primary and secondary users.
Specifically, by backscattering the ambient RF signals transmitted by the primary users,
the passive backscatter devices (BDs) in SR systems can modulate their information without using dedicate spectrum or energy \cite{SR1,SR2,SR3,SR4,SR5}.
Note that the BDs in SR systems are predominantly Internet of Things (IoT) devices, whose low-cost nature aligns well with the scalability requirements of UAV swarm.
Therefore, in this paper, we propose a UAV-SR system to achieve spectrum- and energy-efficient communication, where a leading UAV acts as the primary transmitter (PT) to send communication signals to the BS,
while the rest of the nearby UAVs in the swarm act as BDs to modulate their information by passively backscattering the incident signals from the leading UAV.
Since the backscattering link experiences double-fading attenuation, the dense deployment characteristic of UAV swarm makes the energy utilization more efficient,
which in turn brings advantages to SR.

%For numerous SR applications such as healthcare monitoring, environmental monitoring, and vehicle-to-everything (V2X) networks \cite{SR1},
%BDs require concurrent communication and positioning functions, and the integration of sensing and communication within SR systems can optimize spectral and hardware resource efficiency.
%However, research on this topic is still in its infancy and relatively few studies have been conducted.
%To explore the capability of ISAC in SR systems, \cite{SRISAC1} analyzes the communication rates for the user and BD,
%and the sensing rate at the base station (BS).
%For SR systems in mobile scenarios, \cite{SRISAC2} proposes a direction of arrival (DoA)
%estimator to locate the moving BD and DoA-assisted detectors to detect SR symbols.
%In addition, the system sum-rate maximization problem is investigated for a reconfigurable intelligent surface (RIS) assisted SR systems to simultaneously
%provide sensing and communication functionality \cite{SRISAC3}.
%To address the huge overhead challenge of acquiring channel state information (CSI) in ISAC SR systems with multiple BDs,
%\cite{SRISAC4} introduces a off-grid sparse Bayesian learning algorithm for simultaneous sensing and symbol detection.
On the other hand, in order to obtain accurate sensing for UAV swarm cost-effectively, integrated super-resolution sensing and communication (IS$^{2}$AC) can be utilized,
which incorporates super-resolution signal processing algorithms in integrated sensing and communication (ISAC) systems to achieve extremely high
sensing performance for key parameters \cite{zhangchaoyue,ISSACJournal}.
To estimate both the azimuth and elevation angles of UAV swarm for sensing and provide full-dimensional beamforming for communication,
three-dimensional (3D) multiple-input multiple-output (MIMO) becomes a critical technology in such an IS$^{2}$AC based UAV-SR system,
by placing antennas in a two-dimensional (2D) grid at the BS.
However, simply increasing the number of antenna elements inevitably raises hardware costs, signal processing complexity, and power consumption.
To address these challenges, sparse MIMO architectures offer a viable pathway. By eliminating the half-wavelength antenna spacing constraint,
sparse MIMO enables array aperture expansion without additional physical elements or RF chains \cite{lixinrui,wanghuizhi}.
Compared to conventional compact MIMO with half-wavelength antenna spacing, sparse MIMO achieves a significant enhancement in degrees of freedom (DoFs)
and is able to identify more sources by deploying the antennas with optimized sparse topologies.
Typical sparse configurations that have been extensively investigated in the literature include uniform sparse array (USA) \cite{wanghuizhi}, minimum redundant array (MRA)
\cite{MRA}, nested array (NA) \cite{NA1}, modular array (MoA) \cite{MoA}, and coprime array (CPA) \cite{CPA}.
In particular, \cite{2Dsparseaarray} reviews existing works on estimating the azimuth angles and elevation angles with 2D sparse arrays,
which can be roughly divided into parallel linear arrays \cite{parallel1,parallel2,parallel3}, non-parallel linear arrays \cite{Lshaped1,Lshaped2,nonparallel1,nonparallel2}, planar arrays \cite{NA2D1,NA2D2,planar}, and other geometries \cite{other1,other2}.

Among these architectures, NAs have gained great attention since they can be systematically devised
and are able to form virtual arrays without holes \cite{NA1,Lshaped1,Lshaped2,NA2D1,NA2D2,minhongqi}.
For 2D nested arrays, \cite{Lshaped1} devises a scheme to find 2D direction for L-shaped array structured by two NAs
by pair-matching of the elevation and azimuth estimated angles.
\cite{Lshaped2} proposes a 2D DoA estimation algorithm via correlation matrix reconstruction for L-shaped nested array (LNA).
\cite{NA2D1} designs a planar nested array (PNA) and further provides the practical algorithms and beamforming in \cite{NA2D2} .
To fully exploit the potential of NAs for ISAC system, \cite{minhongqi} provides the beam pattern
analysis of NAs and studies the beam pattern metrics such as the main lobe beam
width, peak-to-local-minimum ratio, and prominent side lobes height.

Motivated by the above discussions, we consider an IS$^{2}$AC based low-altitude UAV-SR system as shown in Fig. \ref{systemmodel}, with a leading UAV and several low-power UAVs surrounding it.
Furthermore, by deploying 3D sparse MIMO, the BS can achieve accurate sensing and efficient communication for UAV swarm.
However, for SR systems with UAV swarm, traditional channel state information (CSI) estimation relying exclusively on pilot training results in significant overhead.
Therefore, we propose a super-resolution sensing assisted channel estimation algorithm.
By exploiting the fact that super-resolution sensing algorithms do not rely on dedicated pilot signals,
we first use the subspace-based super-resolution algorithms to estimate the DoAs of PT and BDs, thereby obtaining the sensing parameters of UAV swarm.
Then, by using very few pilots, the channel path gains of all UAVs can be obtained where
the beamforming gain can be achieved with the previously estimated DoAs.
Our main contributions are summarized as follows:

\begin{itemize}
	\item First, we provide rigorous mathematical model for the IS$^{2}$AC SR system for low-altitude UAV swarm with 3D sparse MIMO,
where both the LNA and PNA are considered at the BS.
The communication performance is then analyzed by deriving the achievable communication rate expressions of the UAV swarm.
Furthermore, to evaluate the sensing performance, we derive the beam patterns for LNA, PNA and the conventional compact uniform planar array (UPA).
We show that for the same number of physical antenna elements, the main lobe beam width (BW) of LNA is narrower than that of PNA and UPA, which brings greater spatial resolution, but also denser grating lobes.
Prominent side lobe height (SLH) for LNA and PNA are further analyzed.

\item Next, we propose the super-resolution sensing assisted channel estimation method, by exploiting the fact that subspace-based super-resolution sensing algorithms do not require dedicated pilot signals.
Specifically, by constructing virtual arrays of LNA and PNA to form a larger aperture compared with the conventional compact array, the elevation and azimuth angles of UAV swarm are accurately estimated with super-resolution sensing algorithms. Then, by matching the angles to obtain beamforming gain, channel multi-path gains are
estimated accurately with little pilots for sparse MIMO.

\item Furthermore, to evaluate the sensing-assisted channel estimation performance, we analyze the mean square error (MSE) of the estimation of channel gains of UAV swarm.
It is demonstrated that our proposed IS$^{2}$AC-based channel estimation method can achieve better estimation performance than the benchmark scheme.
By comparing the MSE of the estimation of channel gains and sum rates of BDs for LNA, PNA, and UPA, numerical results are provided to validate our theoretical analysis.

\end{itemize}

The rest of this paper is organized as follows. Section II
presents the system model of IS$^{2}$AC based SR with low-altitude UAV swarm.
Section III introduces the sparse array architectures of LNA and PNA.
Section IV derives the communication rates for UAV swarm and
analyzes the beam patterns for LNA, PNA, and UPA.
Section V proposes the IS$^{2}$AC based channel estimation method for these two sparse arrays
by estimating 2D angles first, followed by the estimation of multi-path channel gain coefficients.
In Section VI, numerical results are presented to
validate our theoretical studies. Finally, we conclude the paper
in Section VII.

\emph{Notations:} In this paper, italic letters denote scalars.
The boldface lower- and upper-case letters denote vectors and matrices, respectively.
$\mathbf{a}^{\mathrm{T}}$, $\mathbf{a}^{\mathrm{H}}$ and $\| \mathbf{a}\|$
give the transpose, Hermitian transpose, and Euclidean norm of a vector $\mathbf{a}$, respectively.
${{\mathbf{A}}^{*}}$, ${{\mathbf{A}}^{\mathrm{T}}}$ and ${{\mathbf{A}}^{\mathrm{H}}}$ denote the conjugate, transpose and Hermitian transpose of a matrix $\mathbf{A}$, respectively.
$\mathrm{vec}(\mathbf{A})$ represents stacking the columns of matrix $\mathbf{A}$ into a column vector.
${\rm tr}(\mathbf{B})$ denotes the trace of a square matrix $\mathbf{B}$.
$\otimes$ refers to the Kronecker product.
${\mathbf{I}_{M}}$ is an ${M\times M}$ identity matrix.
$\mathbb{C}^{M\times N}$ denotes the space of ${M\times N}$ matrices with complex entries.
$\mathbb{E}_{X}[\cdot]$ is the expectation taken over the random variable $X$.
$\mathcal{CN}(\mathbf{x},\mathbf{\Sigma })$ denotes the
distribution of a circularly symmetric complex Gaussian (CSCG) variable with mean $\mathbf{x}$ and variance $\mathbf{\Sigma }$.
\section{System Model}
\begin{figure}[!t]
  \centering
  \centerline{\includegraphics[width=2.9in,height=2.9in]{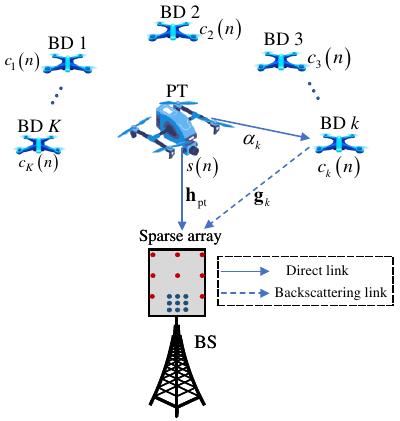}}
  \caption{IS$^{2}$AC-SR systems for low-altitude UAV swarm with 3D sparse MIMO.}
  \label{systemmodel}
  \vspace{-0.3cm}
  \end{figure}
%\begin{figure}[htbp]%figure是双栏单列 figure*单栏一整列
%%\setlength{\abovecaptionskip}{-0.1cm}
%% \setlength{\belowcaptionskip}{-0.1cm}
%  \centering
%    \begin{subfigure}{0.5\textwidth}
%      \centering
%      \includegraphics[width=1\linewidth]{NASRmodel.pdf}
%        \caption{The BS is equipped with L-shaped NA, where two 1D NAs deploy along $x$ and $z$ axes, respectively.}
%        \label{hybridgammavsM}
%    \end{subfigure}   %      \hfill  % 这个\hfill指令为插入弹性长度的空白，看情况选择加不加。
%     \begin{subfigure}{0.5\textwidth}
%      \centering
%      \includegraphics[width=1\linewidth]{2DNASRmodel.pdf}
%        \caption{The BS is equipped with PNA, where the 2D NA is deployed in the $y$-$z$ plane.}
%        \label{hybridgammavsPd}
%    \end{subfigure}
%\caption{IS$^{2}$AC-SR systems with $K$ BDs and one BS.}\label{systemmodel}
%% \vspace{-0.1cm}
%\end{figure}
As shown in Fig. \ref{systemmodel}, we consider the uplink IS$^{2}$AC-SR
systems with one BS and a low-altitude UAV swarm, where the latter consists of one leading active UAV and $K$ passive UAVs. UAV swarm is equipped with one antenna, and the BS is equipped with $M$-element sparse array.
The leading UAV acts as PT to transmit communication signals to the BS,
while the $K$ passive UAVs act as BDs to modulate their information by backscattering the RF signals from the PT without consuming additional spectrum or power.
Let ${{\mathbf{h}}_{\rm{pt}}}\in {{\mathbb{C}}^{M\times 1}}$ denote the direct-link channel from PT to the BS, ${\alpha}_{k}$ denote the channel from PT to BD $k$,
and ${{\mathbf{g}}_{k}}\in {{\mathbb{C}}^{M\times 1}}$ denote the channel from BD $k$ to the BS.
Then the corresponding cascaded channel from the PT to the BS via BD $k$ is denoted as
${\alpha}_{k}{{\mathbf{g}}_{K}}\in {{\mathbb{C}}^{M\times 1}}$.
It is assumed that the direct link channel from the PT to the BS and the backscatter link channels from the PT to the BS via BDs are line-of-sight (LoS) channels.
Let ${{\gamma }_{\mathrm{pt}}}$ denote the complex amplitude from PT to the BS,
$\mathbf{a}\left( {{\theta }_{\mathrm{pt}}},{{\phi }_{\mathrm{pt}}} \right)$ denote the array response vector at the BS associated with the PT, with ${{\theta }_{\mathrm{pt}}}$ and ${{\phi }_{\mathrm{pt}}}$ being the elevation and azimuth angles at the BS, respectively.
Then ${{\mathbf{h}}_{\rm{pt}}}$ can be expressed as
\begin{equation}
{{{\mathbf{h}}_{\mathrm{pt}}}={{\gamma }_{\mathrm{pt}}}\mathbf{a}\left( {{\theta }_{\mathrm{pt}}},{{\phi }_{\mathrm{pt}}} \right)}.
\end{equation}
Further denote ${{\beta }_{k}}$ as the complex amplitude from BD $k$ to the BS,
$\mathbf{a}\left( {{\theta }_{k}},{{\phi }_{k}} \right)$ as the array response vector at the BS associated with $k$-th BD, with ${{\theta }_{k}}$ and ${{\phi }_{k}}$ being the elevation and azimuth angles at the BS.
${{\mathbf{g}}_{k}}$ can be expressed as
\begin{equation}
{{{\mathbf{g}}_{k}}={{\beta }_{k}}\mathbf{a}\left( {{\theta }_{k}},{{\phi }_{k}} \right)}.
\end{equation}

We consider the setup of parasite SR \cite{R.Long}, where the symbol rate of the PT is equal to that of the BDs.
Denote the independent and identically distributed (i.i.d.) information-bearing symbols of the PT and BD $k$ as $s(n)$ and $c_k(n)$, respectively, both of which follow circularly symmetric complex
Gaussian (CSCG) distribution with normalized power, i.e., $s(n)\sim\mathcal{C}\mathcal{N}(0, 1)$, and $c_k(n)\sim\mathcal{C}\mathcal{N}(0,1)$, $k=1,\cdots,K$.
Let $P_d$ denote the transmit power by the PT during the data transmission phase.
Thus, the received signal at the BS can be written as
\begin{equation}
\setlength\abovedisplayskip{1pt}
\setlength\belowdisplayskip{1pt}
{\mathbf{y}_{d}\left( n \right)\!=\!\sqrt{P_d}{{\mathbf{h}}_{\mathrm{pt}}}s\left( n \right)\!+\!\sum\limits_{k=1}^{K}\sqrt{P_d}{\alpha}_{k}{{\mathbf{g}}_{k}}{{c}_{k}}\left( n \right)s\left( n \right)\!+\!\mathbf{u}_{d}\left( n \right),} \label{originy}
\end{equation}
where $\mathbf{u}_{d}(n)\in \mathbb{C}^{M\times 1}$ denotes the CSCG noise with zero mean and power ${{\sigma }^{2}}$, i.e., $\mathbf{u}_{d}(n)\sim\mathcal{C}\mathcal{N}\left( \mathbf 0, \sigma^2{\mathbf{I}}_{M}\right)$.

\begin{figure}[!t]
  \centering
  \centerline{\includegraphics[width=2.3in,height=0.6in]{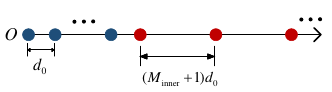}}
  \caption{A typical 1D NA.}
  \label{1DNA}
  \vspace{-0.3cm}
  \end{figure}

To investigate the performance of sparse arrays, we consider NA deployments for the BS.
As shown in Fig. \ref{1DNA}, for a typical one-dimensional (1D) NA with $M_{\mathrm{na}}$ elements \cite{NA1}, it consists of an $M_{\mathrm{inner}}$-element inner subarray with element spacing $d_0=\lambda /2$ and an
$M_{\mathrm{outer}}$-element outer subarray with element spacing $(M_{\mathrm{inner}}+1)d_0$,
where $M_{\mathrm{na}}=M_{\mathrm{inner}}+M_{\mathrm{outer}}$ and $\lambda$ is the wavelength.
Denote the position vector of the $i$th antenna as ${{\mathbf{{x}}}_{i}}$, the difference co-array is defined by the set $\left\{ {{{\mathbf{{x}}}}_{i}}-{{{\mathbf{{x}}}}_{j}} \right\},\forall i,j=1,\cdots ,M_{\mathrm{na}}$.
However, for low-altitude UAV swarm which involves both azimuth and elevation angles, 1D nested array is insufficient. Therefore, in the following, 2D NA deployments with different complexity and degrees of freedom (DoF) are specifically introduced.
\section{2D Nested Array Configuration}
In order to resolve both azimuth and elevation angles for low-altitude UAVs, the LNA and PNA are deployed at the BS to form the virtual array without holes.
LNA is the most straightforward extension from 1D NA to 2D NA, since it is composed of two orthogonal NAs, each of which can form a 1D virtual array without holes.
On the other hand, PNA is a more general 2D sparse NA that can form a 2D virtual array without holes.

\begin{figure}[!t]
  \centering
  \centerline{\includegraphics[width=2.5in,height=2.2in]{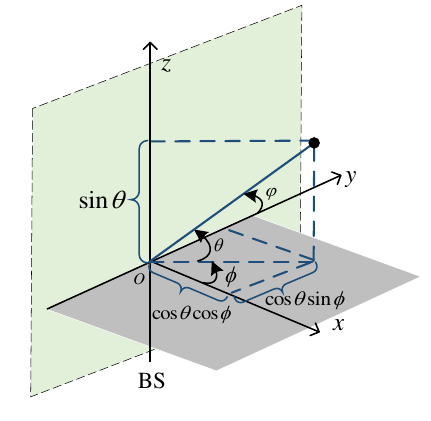}}
  \caption{Illustration of the 3D local coordinate system for the BS and the corresponding spatial angles.}
  \label{3Dzuobiao}
  \vspace{-0.3cm}
  \end{figure}
As shown in Fig. \ref{3Dzuobiao}, in the 3D local coordinate system,
the wave propagation direction of one channel path received by the BS projected onto the $x$, $y$, and $z$ axes are $\cos \theta \cos \phi$, $\cos \theta \sin \phi $ and $\sin \theta $, respectively.
For 2D NA located in the $y$-$z$ plane with $M$ antennas, the element position can be defined as $\left( {{y}_{m}},{{z}_{m}} \right), m=1, 2, ..., M$.
Then the difference of the signal propagation distance
between the array position $\left( {{y}_{m}},{{z}_{m}} \right)$ and $O$ at the BS for the direct link from PT can be expressed as
${{d}_{\mathrm{pt},m,n}}=\cos {{\theta }_{\mathrm{pt}}}\sin {{\phi }_{\mathrm{pt}}}{{y}_{m}}+\sin {{\theta }_{\mathrm{pt}}}{{z}_{m}}$, and that for the backscatter link from $k$-th BD is expressed as
${{d}_{k,m,n}}=\cos {{\theta }_{k}}\sin {{\phi }_{k}}{{y}_{m}}+\sin {{\theta }_{k}}{{z}_{m}}$, $m=1,...,M$, $k=1,...,K$, respectively.
For convenience of notation, we can further define the incident direction ${{\varphi }}$ depicted in Fig. \ref{3Dzuobiao} so that $\cos \varphi =\cos {{\theta }}\sin {{\phi }}$.
\subsection{LNA}
\begin{figure}[!t]
  \centering
  \centerline{\includegraphics[width=1.7in,height=1.5in]{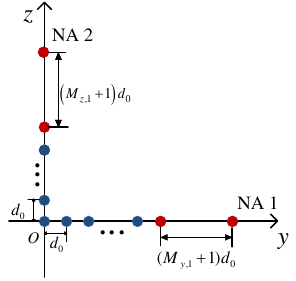}}
  \caption{An illustration of the LNA, which consists of two 1D NAs deployed along $y$ and $z$ axes, respectively. The inner and outer ULAs are respectively denoted in blue and red.}
  \label{LshapedNA}
  \vspace{-0.3cm}
  \end{figure}
First, to gain some insights, we consider the LNA for its appealing characteristic of easy deployment.
As shown in Fig. \ref{LshapedNA}, the BS is equipped with an LNA deployed in the $y$-$z$ plane.
The L-shaped array is constructed by two double-level NAs along the $y$ and $z$ axes with $M_y=M_{y,1}+M_{y,2}$ and $M_z=M_{z,1}+M_{z,2}$ elements,
respectively.
Specifically, the inner compact uniform linear arrays (ULA) have $M_{y,1}$ and $M_{z,1}$ elements with spacing $d_0=\lambda /2$, while the outer sparse ULA has $M_{y,2}$ and $M_{z,2}$ elements with spacing $(M_{y,1}+1)d_0$ and $(M_{z,1}+1)d_0$, respectively.
A reference point $O$ is set at the origin of the coordinate system.
Since two NAs may share the element at the origin, we have $M=M_y+M_z-1$.
%Define the element position in the $y$-$z$ plane as $\left( {{y}_{m}},{{z}_{n}} \right), m=1, 2, ..., M_y, n=1, 2, ..., M_z$.
%The wave propagation direction vector for the direct link from PT and the backscatter link from $k$-th BD
%can be expressed as $\left( \cos {{\theta }_{\text{pt}}}\sin {{\phi }_{\text{pt}}},\sin {{\theta }_{\text{pt}}} \right)$ and
%$\left( \cos {{\theta }_{k}}\sin {{\phi }_{k}},\sin {{\theta }_{k}} \right)$, respectively.
%Then the difference of the signal propagation distance
%between the array position $\left( {{y}_{m}},{{z}_{n}} \right)$ and $O$ at the BS for the direct link from PT can be expressed as
%${{d}_{\mathrm{pt},m,n}}=\cos {{\theta }_{\mathrm{pt}}}\sin {{\phi }_{\mathrm{pt}}}{{y}_{m}}+\sin {{\theta }_{\mathrm{pt}}}{{z}_{n}}$, and that for the backscatter link from $k$-th BD is expressed as
%${{d}_{k,m,n}}=\cos {{\theta }_{k}}\sin {{\phi }_{k}}{{y}_{m}}+\sin {{\theta }_{k}}{{z}_{n}}$, $m=1,...,M_y, n=1, ..., M_z$, $k=1,...,K$, respectively.
%For convenience of notation, we define the incident direction ${{\varphi }}$ depicted in Fig. \ref{LshapedNA} so that $\cos \varphi =\cos {{\theta }}\sin {{\phi }}$.

Since the layouts of the two NAs at the $y$-axis and $z$-axis are similar, we take the antenna locations of $z$ axes
as an example.
By dividing by ${d}_{0}$, they can be normalized expressed by the integer set $\mathcal{D}_{z}=\left\{ {{{\bar{d}}}_{z,1}},{{{\bar{d}}}_{z,2}},\cdots ,{{{\bar{d}}}_{z,M_z}} \right\}$, where ${{\bar{d}}_{z,n}}=\frac{{{d}_{z,n}}}{{{d}_{0}}}$
with ${d}_{z,n}$ denoting the distance between the origin $O$ and the $n$-th element at the $z$ axes.

\emph{Definition 1}: The NA is defined according to the integer
parameter pair $(M_{z,1}, M_{z,2})$, which is the union of an inner compact ULA
and an outer sparse ULA, i.e., $\mathcal{D}_{z}={{\mathcal{D}}_{z,in}}\cup {{\mathcal{D}}_{z,ou}}$ \cite{minhongqi}, where
\begin{equation}
\begin{split}
\setlength\abovedisplayskip{1pt}
\setlength\belowdisplayskip{1pt}
 \text{Compact } & \text{ULA } {\mathcal{D}_{z,in}}:\left\{ 0,1,\cdots ,{{M}_{z,1}-1} \right\}, \\
 \text{Sparse } &  \text{ULA } {\mathcal{D}_{z,ou}}:\\
 & \left\{ {{M}_{z,1}},2\left( {{M}_{z,1}}+1 \right)-1,\cdots ,{{M}_{z,2}}\left( {{M}_{z,1}}+1 \right)-1 \right\}.
\end{split}
\end{equation}
The difference co-array for antennas on the $z$ axis is defined as the set ${\mathcal{D}_{\mathrm{vir},z}}=\left\{ {{d}_{z,i}}-{{d}_{z,j}},i,j=1,2,\cdots ,M_z \right\}$.
%Then the antenna locations can be expressed as an integer set $S=\left\{ m{{d}_{0}},m=1,\cdots ,{{M}_{1}},n\left( {{M}_{1}}+1 \right){{d}_{0}},n=1,\cdots ,{{M}_{2}} \right\}\triangleq \left\{ {{d}_{1}},{{d}_{2}},\cdots ,{{d}_{M}} \right\}$ \cite{NA}.
For LNA, the element locations in the $y$-$z$ plane can be expressed as $\left( {{\bar{d}}_{y,m}{d}_{0}},0 \right), m=1, 2, ..., M_y$
with ${\bar{d}}_{y,m}$ being the normalized antenna locations at the $y$ axes, and $\left( 0,{{\bar{d}}_{z,n}{d}_{0}} \right), n=1, 2, ..., M_z$.
The array response of NA 1 and NA 2 associated with the PT can thus be respectively written as
${{\mathbf{a}}_{y}}\left( {{\varphi }_{\mathrm{pt}}} \right)={{\big[ 1,{{e}^{j\pi {{{\bar{d}}}_{y,2}}\cos {{\varphi }_{\mathrm{pt}}}}},\cdots ,{{e}^{j\pi {{{\bar{d}}}_{y,M_y}}\cos {{\varphi }_{\mathrm{pt}}}}} \big]}^{\mathrm{T}}}$,
${{\mathbf{a}}_{z}}\left( {{\theta }_{\mathrm{pt}}} \right)={{\big[ 1,{{e}^{j\pi {{{\bar{d}}}_{z,2}}\sin {{\theta }_{\mathrm{pt}}}}},\cdots ,{{e}^{j\pi {{{\bar{d}}}_{z,M_z}}\sin {{\theta }_{\mathrm{pt}}}}} \big]}^{\mathrm{T}}}$.
By combining the array response of NA 1 and NA 2 and removing the duplicate entry corresponding to the element at the origin $O$ for NA 2,
the array response vector of the LNA associated with the PT can be obtained as $\mathbf{a}\left( {\theta }_{\mathrm{pt}},{{\varphi }_{\mathrm{pt}}} \right)=\big[{{\mathbf{a}}^{\mathrm{T}}_{y}}\left( {{\varphi }_{\mathrm{pt}}} \right),{{\mathbf{a}}^{\mathrm{T}}_{z,\mathrm{de}}}\left( {{\theta }_{\mathrm{pt}}} \right)\big]^{\mathrm{T}}$, where ${{\mathbf{a}}_{z,\mathrm{de}}}\left( {{\theta }_{\mathrm{pt}}} \right)={{\big[ {{e}^{j\pi {{{\bar{d}}}_{z,2}}\sin {{\theta }_{\mathrm{pt}}}}},\cdots ,{{e}^{j\pi {{{\bar{d}}}_{z,M_z}}\sin {{\theta }_{\mathrm{pt}}}}} \big]}^{\mathrm{T}}}$.
Furthermore, $\mathbf{a}\left( {\theta }_{k},{{\varphi }_{k}} \right)$ can be obtained with a similar form as $\mathbf{a}\left( {\theta }_{k},{{\varphi }_{k}} \right)=\big[{{\mathbf{a}}^{\mathrm{T}}_{y}}\left( {{\varphi }_{k}} \right),{{\mathbf{a}}^{\mathrm{T}}_{z,\mathrm{de}}}\left( {{\theta }_{k}} \right)\big]^{\mathrm{T}}$,
where
${{\mathbf{a}}_{y}}\left( {{\varphi }_{k}} \right)={{\big[ 1,{{e}^{j\pi {{{\bar{d}}}_{y,2}}\cos {{\varphi }_{k}}}},\cdots ,{{e}^{j\pi {{{\bar{d}}}_{y,M_y}}\cos {{\varphi }_{k}}}} \big]}^{\mathrm{T}}}$,
${{\mathbf{a}}_{z,\mathrm{de}}}\left( {{\theta }_{k}} \right)={{\big[{{e}^{j\pi {{{\bar{d}}}_{z,2}}\sin {{\theta }_{k}}}},\cdots ,{{e}^{j\pi {{{\bar{d}}}_{z,M_z}}\sin {{\theta }_{k}}}} \big]}^{\mathrm{T}}}$.
It can be found that the array response of NA 1 ${{\mathbf{a}}_{y}}\left( {{\varphi }_{\mathrm{pt}}} \right)$ and ${{\mathbf{a}}_{y}}\left( {{\varphi }_{k}} \right)$ are only related to $\varphi$, while
the array response of NA 2 ${{\mathbf{a}}_{z,\mathrm{de}}}\left( {{\theta }_{\mathrm{pt}}} \right)$ and ${{\mathbf{a}}_{z,\mathrm{de}}}\left( {{\theta }_{k}} \right)$ are only related to $\theta$.
%\begin{small}
%\begin{equation}
%\begin{split}
%& \mathbf{a}\left( {{\theta }_{\mathrm{ue}}},{{\phi }_{\mathrm{ue}}} \right)\\
%& \!=\!{{\big[ 1,{{e}^{j\pi {{{\bar{d}}}_{y,2}}\cos {{\phi }_{\mathrm{ue}}}}},\!\cdots\! ,{{e}^{j\pi {{{\bar{d}}}_{y,M_y}}\cos {{\phi }_{\mathrm{ue}}}}},{{e}^{j\pi {{{\bar{d}}}_{z,2}}\sin {{\theta }_{\mathrm{ue}}}}},\!\cdots \!,{{e}^{j\pi {{{\bar{d}}}_{z,M_z}}\sin {{\theta }_{\mathrm{ue}}}}} \big]}^{\mathrm{T}}}.
%\end{split}
%\end{equation}
%\end{small}

%$\mathbf{a}\left( {\theta }_{k},{{\phi }_{k}} \right)={{\big[ 1,{{e}^{j\pi {{{\bar{d}}}_{2}}\cos {{\theta }_{k}}}},\cdots ,{{e}^{j\pi {{{\bar{d}}}_{M}}\cos {{\theta }_{k}}}},{{e}^{j\pi {{{\bar{d}}}_{2}}\cos {{\phi }_{\mathrm{ue}}}}},\\ \cdots ,{{e}^{j\pi {{{\bar{d}}}_{M}}\cos {{\phi }_{k}}}} \big]}^{\mathrm{T}}}\in {{\mathbb{C}}^{\left( 2M-1 \right)\times 1}}.$

\subsection{PNA}
\begin{figure}[htbp]%figure是双栏单列 figure*单栏一整列
  \centering
    \begin{subfigure}{0.2\textwidth}
      \centering
      \includegraphics[width=\linewidth]{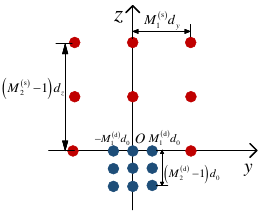}
        \caption{A PNA with the locations of physical array given in Configuration II in \cite{NA2D1}.}
        \label{fenjie1}
    \end{subfigure}   %      \hfill  % 这个\hfill指令为插入弹性长度的空白，看情况选择加不加。
     \begin{subfigure}{0.2\textwidth}
      \centering
      \includegraphics[width=\linewidth]{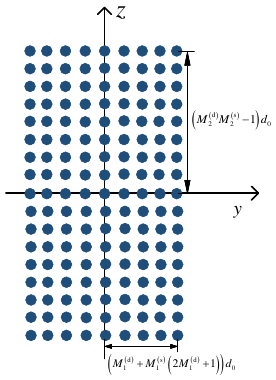}
        \caption{The resulting difference co-array, where the positive and negative halves form a continuum.}
        \label{fenjie2}
    \end{subfigure}
    \caption{An illustration of the PNA in the $y$-$z$ plane, where the compact and sparse UPAs are respectively denoted in blue and red.}
    \label{PNA}
      \vspace{-0.3cm}
      \end{figure}

%\begin{figure}[!t]
%  \centering
%  \centerline{\includegraphics[width=3.3in,height=2.9in]{PNA.pdf}}
%  \caption{The BS is equipped with PNA, where the 2D NA is deployed in the $y$-$z$ plane.}
%  \label{PNA}
%%  \vspace{-0.3cm}
%  \end{figure}
Next, we consider the more general PNA.
As shown in Fig. \ref{PNA}, we consider the Antenna Configuration II in \cite{NA2D1}, which is a general antenna configuration strategy where the difference co-array contains contiguous elements that cannot be satisfied by Antenna Configuration I in \cite{NA2D1}.
Specifically, a PNA is composed by a conventional compact UPA with ${{M}^{\left( \mathrm{d} \right)}}$ elements and a sparse UPA with ${{M}^{\left( \mathrm{s} \right)}}$ elements, and it is characterized by four parameters: $M_{1}^{\left( \mathrm{d} \right)}$, $M_{2}^{\left( \mathrm{d} \right)}$,
$M_{1}^{\left( \mathrm{s} \right)}$, and $M_{2}^{\left( \mathrm{s} \right)}$. The locations of the physical array are illustrated in Fig. \ref{PNA} (a),
where the sparse and compact UPA are placed on opposite sides with respect to the origin.
The conventional compact UPA with ${{M}^{\left( \mathrm{d} \right)}}=\big( 2M_{1}^{\left( \mathrm{d} \right)}+1 \big)M_{2}^{\left( \mathrm{d} \right)}$ elements are deployed at $-M_{1}^{\left( \mathrm{d} \right)}{d}_{0}\le y\le M_{1}^{\left( \mathrm{d} \right)}{d}_{0}$, $\big(-M_{2}^{\left( \mathrm{d} \right)}+1\big){d}_{0}\le z\le 0$ with spacing ${{d}_{0}}=\frac{\lambda }{2}$.
Meanwhile, the sparse UPA has ${{M}^{\left( \mathrm{s} \right)}}=\big( 2M_{1}^{\left( \mathrm{s} \right)}+1 \big)M_{2}^{\left( \mathrm{s} \right)}$ elements located at $-M_{1}^{\left( \mathrm{s} \right)}{d}_{y}\le y\le M_{1}^{\left( \mathrm{s} \right)}{d}_{y}$, $0\le z\le \big(M_{2}^{\left( \mathrm{s} \right)}-1\big){d}_{z}$ with spacing ${{d}_{y}}=\big( 2M_{1}^{\left( \mathrm{d} \right)}+1 \big)d_0$ at $y$ axis and ${{d}_{z}}=M_{2}^{\left( \mathrm{d} \right)}d_0$ at $z$ axis.
Since the compact and sparse UPAs may share the element at the origin,
we have $M={{M}^{\left( \mathrm{d} \right)}}+{{M}^{\left( \mathrm{s} \right)}}-1$.

For Fig. \ref{PNA}, $M_{1}^{\left( \mathrm{d} \right)}=1$, $M_{2}^{\left( \mathrm{d} \right)}=3$, $M_{1}^{\left( \mathrm{s} \right)}=1$, and $M_{2}^{\left( \mathrm{s} \right)}=3$.
Denote the position vector of the $i$-th
antenna of sparse UPA as ${{\mathbf{{a}}}_{s,i}=(y_{s,i},z_{s,i})},1\le i\le {{M}^{\left( \mathrm{s} \right)}}$, and that of the $j$-th antenna of compact UPA as ${{\mathbf{{a}}}_{d,j}=(y_{d,j},z_{d,j})},1\le j\le {{M}^{\left( \mathrm{d} \right)}}$.
The difference co-array is defined as the set $\mathcal{D}_{\mathrm{vir,PNA}}= \pm \left\{ {{{\mathbf{{a}}}}_{s,i}}-{{{\mathbf{{a}}}}_{d,j}},1\le i\le {{M}^{\left( \mathrm{s} \right)}},1\le j\le {{M}^{\left( \mathrm{d} \right)}} \right\}$,
which is shown in Fig. \ref{PNA} (b).
Since the sparse and compact UPAs are deployed on opposite sides with respect to the origin,
the resulting positive and negative halves of the difference co-array contain contiguous elements
and they overlap only along a line on the $y$ axis \cite{NA2D1}.

Denote the array response of PNA associated with the PT as $\mathbf{a}\left( {{\theta }_{\mathrm{pt}}},{{\phi }_{\mathrm{pt}}} \right)=\left[ \begin{matrix}
   {\mathbf{a}_{\mathrm{s}}}\left( {{\theta }_{\mathrm{pt}}},{{\phi }_{\mathrm{pt}}} \right)  \\
   {\mathbf{a}_{\mathrm{d}}}\left( {{\theta }_{\mathrm{pt}}},{{\phi }_{\mathrm{pt}}} \right)  \\
\end{matrix} \right]\in \mathbb{C}^{M\times 1}$,
where ${\mathbf{a}_{\mathrm{s}}}\left( {{\theta }_{\mathrm{pt}}},{{\phi }_{\mathrm{pt}}} \right)\in \mathbb{C}^{{{M}^{\left( \mathrm{s} \right)}}\times 1}$ is the array response of sparse UPA,
${\mathbf{a}_{\mathrm{d}}}\left( {{\theta }_{\mathrm{pt}}},{{\phi }_{\mathrm{pt}}} \right)\in \mathbb{C}^{{\left({M}^{\left( \mathrm{d} \right)}-1\right)}\times 1}$ is the array response of compact UPA except for the duplicate coordinate $(0,0)$ with sparse UPA.
Since the difference of the signal propagation distance between the array position $\left( {{y}_{m}},{{z}_{m}} \right)$ and $O$ at the BS for the direct link from PT are expressed as
${{d}_{\mathrm{pt},m,n}}=\cos {{\theta }_{\mathrm{pt}}}\sin {{\phi }_{\mathrm{pt}}}{{y}_{m}}+\sin {{\theta }_{\mathrm{pt}}}{{z}_{m}}, m=1,...,M$,
the $i$-th element of ${\mathbf{a}_{\mathrm{s}}}\left( {{\theta }_{\mathrm{pt}}},{{\phi }_{\mathrm{pt}}} \right)$ and the $m$-th element of ${\mathbf{a}_{\mathrm{d}}}\left( {{\theta }_{\mathrm{pt}}},{{\phi }_{\mathrm{pt}}} \right)$ can be respectively written as
\begin{equation}
{{{\left[ {\mathbf{a}_{\mathrm{s}}}\left( {{\theta }_{\mathrm{pt}}},{{\phi }_{\mathrm{pt}}} \right) \right]}_{i}}={{e}^{j\pi \left( \cos {{\theta }_{\mathrm{pt}}}\sin {{\phi }_{\mathrm{pt}}}{{y}_{i}}+\sin {{\theta }_{\mathrm{pt}}}{{z}_{i}} \right)}}}
\end{equation}
with $\left\{ \left( {{y}_{i}},{{z}_{i}} \right) \right\}_{i=1}^{{{M}^{\left( \mathrm{s} \right)}}}$ being the coordinates of the sparse UPA,
and
\begin{equation}
{{{\left[ {\mathbf{a}_{\mathrm{d}}}\left( {{\theta }_{\mathrm{pt}}},{{\phi }_{\mathrm{pt}}} \right) \right]}_{m}}={{e}^{j\pi \left( \cos {{\theta }_{\mathrm{pt}}}\sin {{\phi }_{\mathrm{pt}}}{{y}_{m}}+\sin {{\theta }_{\mathrm{pt}}}{{z}_{m}} \right)}}}
\end{equation}
with $\left\{ \left( {{y}_{m}},{{z}_{m}} \right) \right\}_{m=1}^{{{M}^{\left( \mathrm{d} \right)}}}$
being the coordinates of the compact UPA.
Compared with LNA, the elements of the array response for PNA are all related to two spatial angles,
while the elements of that for LNA can be denoted by only one spatial angle $\theta$ or $\varphi$.
This is because LNA is deployed on the coordinate axis, while PNA is deployed on the $y$-$z$ plane.
\section{Analysis of Communication and Sensing Performance}
Next, we analyze the communication and sensing performance for the considered 3D sparse MIMO SR system.
\subsection{Achievable Rate Analysis}
To decode the primary signal $s\left( n \right)$ from the received signal $\mathbf{y}_{d}\left( n \right)$ in \eqref{originy}, the BS applies a receive
beamforming vector ${{\mathbf{w}}_{\mathrm{pt}}}$ with ${{\left\| {{\mathbf{w}}_{\mathrm{pt}}} \right\|}^{2}}=1$, which gives
\begin{small}
\begin{equation}
\begin{aligned}
\setlength\abovedisplayskip{1pt}
\setlength\belowdisplayskip{1pt}
  & {{y}_{\mathrm{pt}}}\left( n \right)=\sqrt{{{P}_{d}}}\mathbf{w}_{\mathrm{pt}}^{\mathrm{H}}{{\mathbf{h}}_{\mathrm{pt}}}s\left( n \right)\\
 & +\sqrt{{{P}_{d}}}\mathbf{w}_{\mathrm{pt}}^{\mathrm{H}}\sum\limits_{k=1}^{K}{{{\alpha }_{k}}}{{\mathbf{g}}_{k}}{{c}_{k}}\left( n \right)s\left( n \right)+\mathbf{w}_{\mathrm{pt}}^{\mathrm{H}}{{\mathbf{u}}_{d}}\left( n \right).
\end{aligned}
\end{equation}
\end{small}
Then the resulting SINR for PT is
\begin{small}
\begin{equation}
\setlength\abovedisplayskip{0.5pt}
\setlength\belowdisplayskip{0.5pt}
{{{\gamma }_{\mathrm{pt}}}=\frac{{{P}_{d}}{{\left| \mathbf{w}_{\mathrm{pt}}^{\mathrm{H}}{{\mathbf{h}}_{\mathrm{pt}}} \right|}^{2}}}{{{P}_{d}}\sum\limits_{k=1}^{K}{{{\left| {{\alpha }_{k}} \right|}^{2}}|\mathbf{w}_{\mathrm{pt}}^{\mathrm{H}}{{\mathbf{g}}_{k}}{{|}^{2}}}+{{\sigma }^{2}}}},\label{SINRw}
\end{equation}
\end{small}where the linear MMSE beamforming for PT can be written as
\begin{small}
\begin{equation}
\setlength\abovedisplayskip{1pt}
\setlength\belowdisplayskip{1pt}
{{{\mathbf{w}}_{\mathrm{pt}}}=\frac{{{\Big({{P}_{d}}\sum\limits_{k=1}^{K}{{{\left| {{{{\alpha }}}_{k}} \right|}^{2}}}{{{\mathbf{{g}}}}_{k}}\mathbf{{g}}_{k}^{\mathrm{H}}+{{\sigma }^{2}}{{\mathbf{I}}_{M}}\Big)}^{-1}}\sqrt{{{P}_{d}}}{{{\mathbf{{h}}}}_{\mathrm{pt}}}}{\Big\|{{\Big({{P}_{d}}\sum\limits_{k=1}^{K}{{{\left| {{{{\alpha }}}_{k}} \right|}^{2}}}{{{\mathbf{{g}}}}_{k}}\mathbf{{g}}_{k}^{\mathrm{H}}+{{\sigma }^{2}}{{\mathbf{I}}_{M}}\Big)}^{-1}}\sqrt{{{P}_{d}}}{{{\mathbf{{h}}}}_{\mathrm{pt}}}\Big\|}.}
\end{equation}
\end{small}
Therefore, the achievable communication rate for PT can be written as
\begin{equation}
\setlength\abovedisplayskip{1pt}
\setlength\belowdisplayskip{1pt}
{{{R}_{\mathrm{pt}}}={{\log }_{2}}\Big( 1+{{P}_{d}}\mathbf{h}_{\mathrm{pt}}^{\mathrm{H}}{{\Big({{P}_{d}}\sum\limits_{k=1}^{K}{{{\left| {{\alpha }_{k}} \right|}^{2}}}{{\mathbf{g}}_{k}}\mathbf{g}_{k}^{\mathrm{H}}+{{\sigma }^{2}}{{\mathbf{I}}_{M}}\Big)}^{-1}}{{\mathbf{h}}_{\mathrm{pt}}} \Big)}.
\end{equation}
After decoding $s\left( n \right)$, the primary signal $s\left( n \right)$ can be subtracted from \eqref{originy} before decoding the BD signals, which results
\begin{equation}
\setlength\abovedisplayskip{1pt}
\setlength\belowdisplayskip{1pt}
{\mathbf{\hat{y}}_{d}\left( n \right)=\sqrt{{{P}_{d}}}\sum\limits_{k=1}^{K}{{{\alpha }_{k}}}{{\mathbf{g}}_{k}}{{c}_{k}}\left( n \right)s\left( n \right)+\mathbf{u}_{d}\left( n \right).}\label{subtractY}
\end{equation}
%Note that \eqref{subtractY} can be seen as a single-input-multiple-output (SIMO) multiple
%access channel (MAC), where MMSE receiver with SIC is known to be capacity-achieving \cite{David}.
Then SIC technology can be used to decode signals for the $K$ BDs.
In particular, the $K$ BDs are sorted from strong to weak based on their channel strength ${{\left\| {{{\alpha} }_{k}}{{{\mathbf{g}}}_{k}} \right\|}^{2}}$.
Without loss of generality, assume that ${{\left\| {{{\alpha} }_{1}}{{{\mathbf{g}}}_{1}} \right\|}^{2}}\ge {{\left\| {{{\alpha} }_{2}}{{{\mathbf{g}}}_{2}} \right\|}^{2}}\ge \cdots \ge {{\left\| {{{\alpha} }_{K}}{{{\mathbf{g}}}_{K}} \right\|}^{2}}$, then the SIC decoding order is $1,2,\cdots ,K$.
Take BD $k$ for an example, where the signals for BDs $1,...,k-1$ have already been decoded and partly removed, and those for BDs $k+1,...,K$ are treated as noise.
Denote the beamforming vector for BD $k$ as ${{\mathbf{w}}_{k}}\in {{\mathbb{C}}^{M\times 1}}$. Then the resulting signal can be written as
\begin{small}
\begin{equation}
\begin{split}
\setlength\abovedisplayskip{1pt}
\setlength\belowdisplayskip{1pt}
  {{y}_{k}}\left( n \right)& \!=\!\sqrt{{{P}_{d}}}\mathbf{w}_{k}^{\mathrm{H}}{{\alpha }_{k}}{{\mathbf{g}}_{k}}s\left( n \right){{c}_{k}}\left( n \right)\!+\!\sqrt{{{P}_{d}}}\mathbf{w}_{k}^{\mathrm{H}}\sum\limits_{i=k+1}^{K}{{{\alpha }_{i}}}{{\mathbf{g}}_{i}}s\left( n \right){{c}_{i}}\left( n \right) \\
 & +\mathbf{w}_{k}^{\mathrm{H}}\mathbf{u}_{d}(n).
\end{split}
\end{equation}
\end{small}
The linear MMSE beamforming for BD $k$ is written as
\begin{small}
\begin{equation}
\setlength\abovedisplayskip{1pt}
\setlength\belowdisplayskip{1pt}
{{{\mathbf{w}}_{k}}=\frac{{{\Big( {{P}_{d}}\sum\limits_{i=k+1}^{K}{{{\left| {{{{\alpha }}}_{i}} \right|}^{2}}}{{{\mathbf{{g}}}}_{i}}\mathbf{{g}}_{i}^{\mathrm{H}}+{{\sigma }^{2}}{{\mathbf{I}}_{M}} \Big)}^{-1}}\sqrt{{{P}_{d}}}{{{{\alpha }}}_{k}}{{{\mathbf{{g}}}}_{k}}}{\Big\| {{\Big( {{P}_{d}}\sum\limits_{i=k+1}^{K}{{{\left| {{{{\alpha }}}_{i}} \right|}^{2}}}{{{\mathbf{{g}}}}_{i}}\mathbf{{g}}_{i}^{\mathrm{H}}+{{\sigma }^{2}}{{\mathbf{I}}_{M}} \Big)}^{-1}}\sqrt{{{P}_{d}}}{{{{\alpha }}}_{k}}{{{\mathbf{{g}}}}_{k}} \Big\|}}.
\end{equation}
\end{small}
Since the squared envelope of $s\left( n \right)$ follows an exponential distribution, and its probability density function (PDF) is $f\left( x \right)={{e}^{-x}},x>0$.
Then the corresponding average achievable rate of the $k$-th BD can be written as \cite{SR2}
\begin{small}
\begin{equation}
\begin{split}
\setlength\abovedisplayskip{1pt}
\setlength\belowdisplayskip{1pt}
  & {{R}_{k}}={{\log }_{2}}\bigg( 1+\frac{{{P}_{d}}{{\left| {{\alpha }_{k}} \right|}^{2}}{{\left| \mathbf{w}_{k}^{\mathrm{H}}{{\mathbf{g}}_{k}} \right|}^{2}}{{\left| s\left( n \right) \right|}^{2}}}{{{P}_{d}}\sum\limits_{i=k+1}^{K}{{{\left| {{\alpha }_{i}} \right|}^{2}}}|\mathbf{w}_{k}^{\mathrm{H}}{{\mathbf{g}}_{i}}{{|}^{2}}{{\left| s\left( n \right) \right|}^{2}}+{{\sigma }^{2}}} \bigg) \\
 & =\int_{0}^{+\infty }{{{e}^{-x}}{{\log }_{2}}\bigg( 1+\frac{{{P}_{d}}{{\left| {{\alpha }_{k}} \right|}^{2}}{{\left| \mathbf{w}_{k}^{\mathrm{H}}{{\mathbf{g}}_{k}} \right|}^{2}}x}{{{P}_{d}}\sum\limits_{i=k+1}^{K}{{{\left| {{\alpha }_{i}} \right|}^{2}}}|\mathbf{w}_{k}^{\mathrm{H}}{{\mathbf{g}}_{i}}{{|}^{2}}x+{{\sigma }^{2}}} \bigg)\mathrm{d}x}.
\end{split}
\end{equation}
\end{small}
Therefore, the sum rate of the $K$ BDs can be written as
\begin{equation}
\setlength\abovedisplayskip{1pt}
\setlength\belowdisplayskip{1pt}
{{R}_{\mathrm{BD}}=\sum\limits_{k=1}^{K}{R}_{k}}.\label{RBDsum}
\end{equation}
\subsection{Analysis of Beam Pattern}
Denote ${{\Delta }_{y}}\triangleq \cos {{\varphi }_{i}}-\cos {{\varphi }_{k}}=\cos {{\theta }_{i}}\sin {{\phi }_{i}}-\cos {{\theta }_{k}}\sin {{\phi }_{k}}\in \left[ -2,2 \right]$, ${{\Delta }_{z}}\triangleq \sin {{\theta }_{i}}-\sin {{\theta }_{k}}\in \left[ -2,2 \right]$.

For the conventional compact UPA with half-wavelength antenna spacing, denote the number of elements at $y$-axis and $z$-axis as ${M}_{y,\mathrm{u}}$ and ${M}_{z,\mathrm{u}}$, respectively, with $M={M}_{y,\mathrm{u}}{M}_{z,\mathrm{u}}$.
Its beam pattern can be written as
\begin{small}
\begin{equation}
\begin{split}
  {\rho }_{k,i}^{\mathrm{upa}}& ={{\bigg| \frac{1}{M}\left( {{\mathbf{a}}^{\mathrm{H}}}\left( {{\theta }_{k}},{{\varphi }_{k}} \right)\mathbf{a}\left( {{\theta }_{i}},{{\varphi }_{i}} \right) \right) \bigg|}^{2}} \\
 & ={{\bigg| \frac{1}{{{M}_{y,\mathrm{u}}}}\frac{\sin \left( \frac{\pi }{2}{{M}_{y,\mathrm{u}}}{{\Delta }_{y}} \right)}{\sin \left( \frac{\pi }{2}{{\Delta }_{y}} \right)} \bigg|}^{2}}{{\bigg| \frac{1}{{{M}_{z,\mathrm{u}}}}\frac{\sin \left( \frac{\pi }{2}{{M}_{z,\mathrm{u}}}{{\Delta }_{z}} \right)}{\sin \left( \frac{\pi }{2}{{\Delta }_{z}} \right)} \bigg|}^{2}}\\
 & \triangleq G^{\mathrm{upa}}\left( {{\Delta }_{y}},{{\Delta }_{z}} \right).
\end{split}
\end{equation}
\end{small}

For the PNA considered in Fig. \ref{PNA}, the beam pattern can be written as
\begin{small}
\begin{equation}
\begin{split}
  & {\rho }_{k,i}^{\mathrm{pna}} ={{\bigg| \frac{1}{M}\left( {{\mathbf{a}}^{\mathrm{H}}}\left( {{\theta }_{k}},{{\varphi }_{k}} \right)\mathbf{a}\left( {{\theta }_{i}},{{\varphi }_{i}} \right) \right) \bigg|}^{2}} \\
 & \!=\!\frac{1}{{{M}^{2}}}{{\bigg| {{\zeta }_{1}}\left( {{\Delta }_{y}},{{\Delta }_{z}} \right)\!+\!{{e}^{j\frac{\pi }{2}\big( M_{1}^{\left( \text{d} \right)}+1 \big)\big( M_{2}^{\left( \text{s} \right)}-1 \big){{\Delta }_{z}}}}{{\zeta }_{2}}\left( {{\Delta }_{y}},{{\Delta }_{z}} \right)\!-\!1 \bigg|}^{2}}\\
 & \triangleq G^{\mathrm{pna}}\left( {{\Delta }_{y}},{{\Delta }_{z}} \right),\label{rhoPNA}
\end{split}
\end{equation}
\end{small}where
\begin{small}
\begin{equation}
{{{\zeta }_{1}}\left( {{\Delta }_{y}},{{\Delta }_{z}} \right)=\frac{\sin \big( \frac{\pi }{2}M_{2}^{\left( \text{d} \right)}{{\Delta }_{z}} \big)\sin \big( \frac{\pi }{2}\big( 2M_{1}^{\left( \text{d} \right)}+1 \big){{\Delta }_{y}} \big)}{\sin \left( \frac{\pi }{2}{{\Delta }_{z}} \right)\sin \left( \frac{\pi }{2}{{\Delta }_{y}} \right)}},
\end{equation}
\end{small}
\begin{small}
\begin{equation}
{{{\zeta }_{2}}\left( {{\Delta }_{y}},{{\Delta }_{z}} \right)\!=\!\frac{\sin \big( \frac{\pi }{2}M_{2}^{\left( \text{d} \right)}M_{2}^{\left( \text{s} \right)}{{\Delta }_{z}} \big)\sin \big( \frac{\pi }{2}\big( 2M_{1}^{\left( \text{d} \right)}\!+\!1 \big)\big( 2M_{1}^{\left( \text{s} \right)}\!+\!1 \big){{\Delta }_{y}} \big)}{\sin \big( \frac{\pi }{2}M_{2}^{\left( \text{d} \right)}{{\Delta }_{z}} \big)\sin \big( \frac{\pi }{2}\big( 2M_{1}^{\left( \text{d} \right)}\!+\!1 \big){{\Delta }_{y}} \big)}}.
\end{equation}
\end{small}

For the LNA considered in Fig. \ref{LshapedNA}, the beam pattern can be written as
\begin{small}
\begin{equation}
\begin{split}
  {\rho }_{k,i}^{\mathrm{lna}}& ={{\Big| \frac{1}{M}\Big( {{\mathbf{a}}^{\mathrm{H}}}\left( {{\theta }_{k}},{{\varphi }_{k}} \right)\mathbf{a}\left( {{\theta }_{i}},{{\varphi }_{i}} \right) \Big) \Big|}^{2}} \\
 & =\frac{1}{{{M}^{2}}}{{\left| \mathbf{a}_{y}^{\mathrm{H}}\left( {{\varphi }_{k}} \right){{\mathbf{a}}_{y}}\left( {{\varphi }_{i}} \right)+\mathbf{a}_{z}^{\mathrm{H}}\left( {{\theta }_{k}} \right){{\mathbf{a}}_{z}}\left( {{\theta }_{k}} \right) \right|}^{2}} \\
 & =\frac{1}{{{M}^{2}}}{{\Big| {{\eta }_{y}}+{{e}^{j\frac{\pi }{2}\left( {{M}_{z,1}}-1 \right){{\Delta }_{z}}-j\frac{\pi }{2}\left( {{M}_{y,1}}-1 \right){{\Delta }_{y}}}}{{\eta }_{z}}-1 \Big|}^{2}}\\
 & \triangleq G^{\mathrm{lna}}\left( {{\Delta }_{y}},{{\Delta }_{z}} \right),\label{rhoLS}
\end{split}
\end{equation}
\end{small}where
\begin{small}
\begin{equation}
\begin{split}
{{\eta }_{y}}& =\frac{\sin \left( \frac{\pi }{2}{{M}_{y,1}}{{\Delta }_{y}} \right)}{\sin \left( \frac{\pi }{2}{{\Delta }_{y}} \right)}\\
& +\frac{{{e}^{j\frac{\pi }{2}\left( {{M}_{y,1}}+1 \right){{M}_{y,2}}{{\Delta }_{y}}}}\sin \left( \frac{\pi }{2}\left( {{M}_{y,1}}+1 \right){{M}_{y,2}}{{\Delta }_{y}} \right)}{\sin \left( \frac{\pi }{2}\left( {{M}_{y,1}}\!+\!1 \right){{\Delta }_{y}} \right)},
\end{split}
\end{equation}
\end{small}
\begin{small}
\begin{equation}
\begin{split}
{{\eta }_{z}}& =\frac{\sin \left( \frac{\pi }{2}{{M}_{z,1}}{{\Delta }_{z}} \right)}{\sin \left( \frac{\pi }{2}{{\Delta }_{z}} \right)}\\
& +\frac{{{e}^{j\frac{\pi }{2}\left( {{M}_{z,1}}+1 \right){{M}_{z,2}}{{\Delta }_{z}}}}\sin \left( \frac{\pi }{2}\left( {{M}_{z,1}}+1 \right){{M}_{z,2}}{{\Delta }_{z}} \right)}{\sin \left( \frac{\pi }{2}\left( {{M}_{z,1}}+1 \right){{\Delta }_{z}} \right)}.
\end{split}
\end{equation}
\end{small}

\begin{figure*}[htbp]%figure是双栏单列 figure*单栏一整列
  \centering
    \begin{subfigure}{0.3\textwidth}
      \centering
      \includegraphics[width=\linewidth]{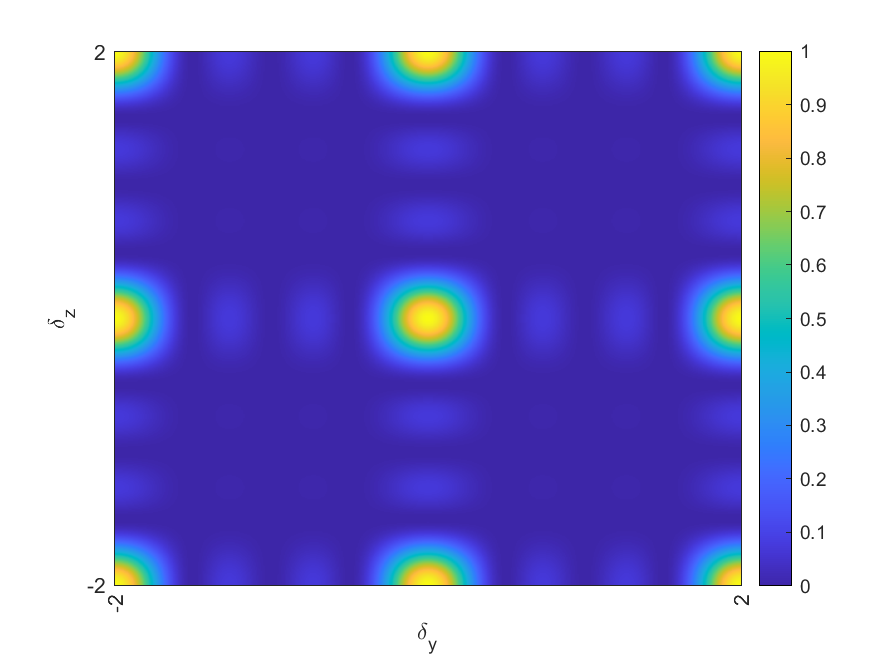}
        \caption{Beam pattern of UPA with ${M}_{y,\mathrm{u}}=4$ and ${M}_{z,\mathrm{u}}=4$.}
        \label{GUPA}
    \end{subfigure}   %      \hfill  % 这个\hfill指令为插入弹性长度的空白，看情况选择加不加。
     \begin{subfigure}{0.3\textwidth}
      \centering
      \includegraphics[width=\linewidth]{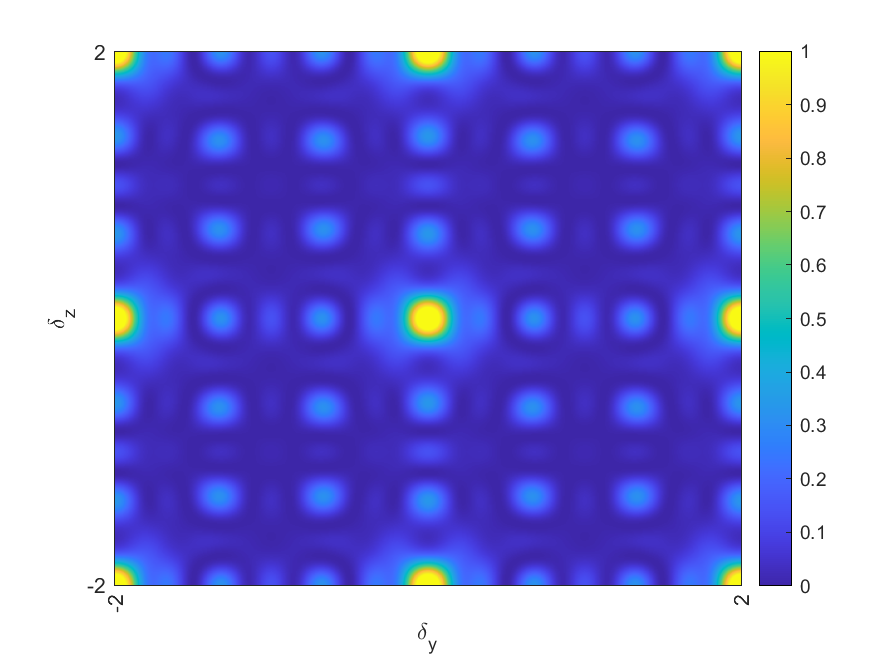}
        \caption{Beam pattern of PNA with ${M}_{1}^{\left(\mathrm{d}\right)}=0$, ${M}_{2}^{\left(\mathrm{d}\right)}=9$, ${M}_{1}^{\left(\mathrm{s}\right)}=3$ and ${M}_{2}^{\left(\mathrm{s}\right)}=1$.}
        \label{GPNA}
    \end{subfigure}
    \begin{subfigure}{0.3\textwidth}
      \centering
      \includegraphics[width=\linewidth]{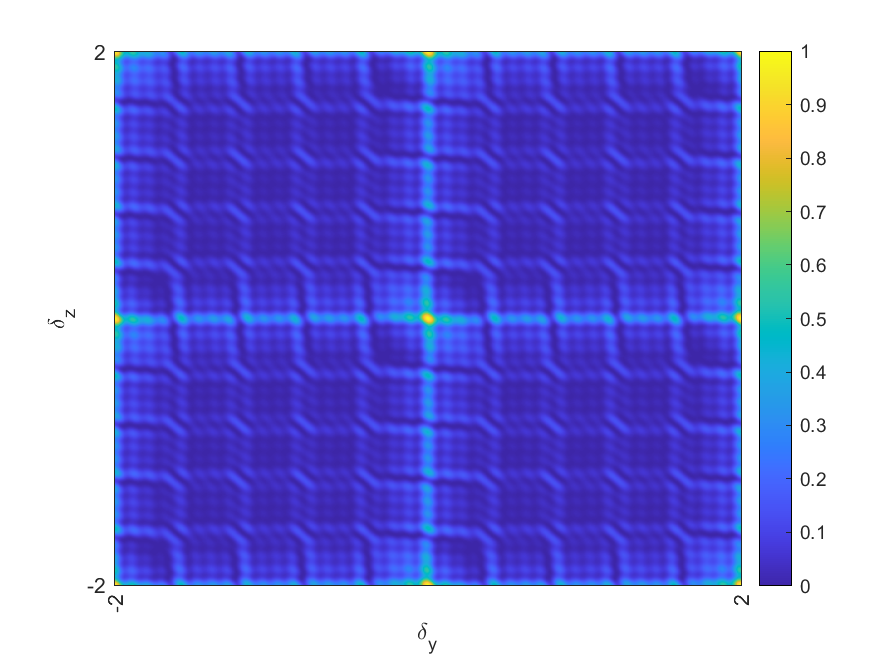}
        \caption{Beam pattern of LNA with ${M}_{y}=8$ and ${M}_{z}=8$.}
        \label{GLSNA}
    \end{subfigure}   %
    \caption{Comparison of beam patterns for UPA with $M=16$, PNA with $M=15$, and LNA with $M=15$.}
     \label{beampattern}
       \vspace{-0.3cm}
      \end{figure*}

In order to obtain an intuitive comparison, Fig. \ref{beampattern} plots the beam pattern for UPA $G^{\mathrm{upa}}\left( {{\Delta }_{y}},{{\Delta }_{z}} \right)$ with $M=16$, PNA $G^{\mathrm{pna}}\left( {{\Delta }_{y}},{{\Delta }_{z}} \right)$ with $M=15$, and LNA $G^{\mathrm{lna}}\left( {{\Delta }_{y}},{{\Delta }_{z}} \right)$ with $M=15$.
%\begin{figure}[!t]
%  \centering
%  \centerline{\includegraphics[width=3.4in,height=2.7in]{GUPA.eps}}
%  \caption{Beam pattern of UPA.}
%  \label{GUPA}
%%  \vspace{-0.2cm}
%  \end{figure}
%\begin{figure}[!t]
%  \centering
%  \centerline{\includegraphics[width=3.5in,height=2.7in]{GPNA.eps}}
%  \caption{Beam pattern of PNA.}
%  \label{GPNA}
%%  \vspace{-0.2cm}
%  \end{figure}
%\begin{figure}[!t]
%  \centering
%  \centerline{\includegraphics[width=3.4in,height=2.7in]{GLSNA.eps}}
%  \caption{Beam pattern of LNA.}
%  \label{GLSNA}
%%  \vspace{-0.2cm}
%  \end{figure}
\subsection{Main lobe Beam Width}
%It is non-trivial to determine the entire 2D main lobe width as in Fig. \ref{GLSNA}, Fig. \ref{GUPA} and Fig. \ref{GPNA}.
As in Fig. \ref{beampattern}, it is non-trivial to determine the null points of the main lobe for all values of $y$ and $z$.
However, by determining the values of the main lobe width region on the $y$-axis and the $z$-axis as $BW_y$ and $BW_z$, the range of the main lobe beam width can be roughly measured.
Note that for LNA and PNA, the main lobe beam width on the $y$-axis and the $z$-axis are defined as $BW_y=2{{\Delta }_{y,\min }}$ and $BW_z=2{{\Delta }_{z,\min }}$, where ${{\Delta }_{y,\min }}$ and ${{\Delta }_{z,\min }}$ are the smallest local minimum points for the positive semi-axis of $y$ and $z$.
%Since it is non-trivial to calculate the 2D BW of $y-z$ plane, we try to respectively measure the BW for $y$ and $z$ axis to characterize the boundaries of the 2D BW.
%The first local minimum point (FLMP) for $y$ and $z$ axis is defined as follows.
%\emph{Definition 1:}
%The FLMP of $G\left( {{\Delta }_{y}},{{\Delta }_{z}} \right)$ of $y$ and $z$ axis are defined as the smallest
%local minimum point on the positive semi-axis, respectively.
%\begin{equation}
%{\Delta _{\min }^{y}=\min \left\{ \Delta |\Delta \in \left\{ \Delta _{i}^{loc,y} \right\}_{i=1}^{n} \right\},}
%\end{equation}
%\begin{equation}
%{\Delta _{\min }^{z}=\min \left\{ \Delta |\Delta \in \left\{ \Delta _{i}^{loc,z} \right\}_{i=1}^{n} \right\},}
%\end{equation}
%where $\left\{ \Delta _{i}^{loc,y} \right\}_{i=1}^{n}$ and $\left\{ \Delta _{i}^{loc,z} \right\}_{i=1}^{n}$ are the $n$ local minimum points of $G\left( {{\Delta }_{y}},{{\Delta }_{z}} \right)$ on the $y$ and $z$ axis, respectively.
%
%The main lobe width of L-shaped NA for $y$ and $z$ axis are defined as $BW_y=2\Delta _{\min }^{y}$ and $BW_z=2\Delta _{\min }^{z}$.
\subsubsection{UPA}
For UPA, the null points of the main lobe on the $y$-axis and the $z$-axis can be obtained by letting $\frac{\pi }{2}{{M}_{y,\mathrm{u}}}{{\Delta }_{y}}=\pm \pi $ and $\frac{\pi }{2}{{M}_{z,\mathrm{u}}}{{\Delta }_{z}}=\pm \pi $, respectively.
Therefore, the beamwidth of the main lobe on the $y$-axis and the $z$-axis are ${{BW}_{y}}=\frac{4}{{{M}_{y,\mathrm{u}}}}$ and ${{BW }_{z}}=\frac{4}{{{M}_{z,\mathrm{u}}}}$, respectively.
\subsubsection{PNA}

\emph{Theorem 1:}
For PNA, the main lobe width on the $y$-axis is bounded by
\begin{small}
\begin{equation}
{\frac{4}{(2M_{1}^{\left( \mathrm{d} \right)}+1)(2M_{1}^{\left( \mathrm{s} \right)}+1)}<BW_y<\frac{4}{2M_{1}^{\left( \mathrm{d} \right)}+1}}.
\end{equation}
\end{small}
Specifically, when $M_{1}^{\left( \text{s} \right)}\ge \frac{M_{2}^{\left( \text{d} \right)}}{M_{2}^{\left( \text{s} \right)}}\big( 2M_{1}^{\left( \text{d} \right)}+1 \big)$,
the main lobe width on the $y$-axis is very close to $\frac{4}{(2M_{1}^{\left( \mathrm{d} \right)}+1)(2M_{1}^{\left( \mathrm{s} \right)}+1)}$ and further upper bounded by
$BW_y<\frac{8}{(2M_{1}^{\left( \mathrm{d} \right)}+1)(2M_{1}^{\left( \mathrm{s} \right)}+1)}$.
\begin{IEEEproof}
Please refer to Appendix A.
\end{IEEEproof}

\emph{Theorem 2:}
The main lobe width on the $z$-axis is bounded by
\begin{equation}
{2\min \left\{ {{\Delta }_{z,2}},{{\Delta }_{z,3}} \right\}<BW_z<\frac{4}{M_{2}^{\left( \mathrm{d} \right)}}},
\end{equation}
where ${{\Delta }_{z,2}}=\frac{2}{M_{2}^{\left( \mathrm{d} \right)}M_{2}^{\left( \mathrm{s}\right)}}$,
${{\Delta }_{z,3}}=\frac{1}{\left( M_{1}^{\left( \text{d} \right)}+1 \right)\left( M_{2}^{\left( \text{s} \right)}-1 \right)}$.
Specifically, when $M_{2}^{\left( \text{s} \right)}\ge \frac{2M_{1}^{\left( \text{d} \right)}+1}{2M_{1}^{\left( \text{s} \right)}+1}\big( 3M_{2}^{\left( \text{d} \right)}-1 \big)$,
the main lobe width on the $z$-axis is very close to $\frac{4}{M_{2}^{\left( \mathrm{d} \right)}M_{2}^{\left( \mathrm{s}\right)}}$ and further upper bounded by
$BW_z<\frac{8}{M_{2}^{\left( \mathrm{d} \right)}M_{2}^{\left( \mathrm{s}\right)}}$.
\begin{IEEEproof}
Please refer to Appendix B.
\end{IEEEproof}
\subsubsection{LNA}
\emph{Theorem 3:}
For LNA, the main lobe width on the $y$-axis is bounded by
\begin{small}
\begin{equation}
{\frac{2}{{\left( M_{y,1}+1 \right)M_{y,2}}}<BW_y<\frac{4}{M_{y,1}}}.
\end{equation}
\end{small}
Specifically, when ${{M}_{y,2}}\ge 3\left( {{M}_{y,1}}+1 \right)$,
the main lobe width on the $y$-axis is very close to and upper bounded by $\frac{4}{\left(M_{y,1}+1\right)M_{y,2}}$.
\begin{IEEEproof}
Please refer to Appendix C.
\end{IEEEproof}

\emph{Theorem 4:}
Following a similar derivation, the main lobe width on the $z$-axis is bounded by
\begin{equation}
{\frac{2}{{\left( M_{z,1}+1 \right)M_{z,2}}}<BW_z<\frac{4}{M_{z,1}}}.
\end{equation}
Furthermore, when ${{M}_{z,2}}\ge 3\left( {{M}_{z,1}}+1 \right)$,
the smallest local minimum points is upper bounded by  $\frac{2}{{\left( M_{z,1}+1 \right)M_{z,2}}}$,
and $BW_z$ is further upper bounded by $BW_z<\frac{4}{\left(M_{z,1}+1\right)M_{z,2}}$.

\subsubsection{Comparison of main lobe width for UPA, PNA and LNA}
Next, we summarize the above derivations to gain some insights.
Without loss of generality, we take the beamwidth of the main lobe on the $y$-axis as an example.
For UPA, the beamwidth of the main lobe on the $y$-axis is expressed as
${BW}_{y,\mathrm{upa}}=\frac{4}{{{M}_{y,\mathrm{u}}}}$.
When $M_{1}^{\left( \text{s} \right)}\ge \frac{M_{2}^{\left( \text{d} \right)}}{M_{2}^{\left( \text{s} \right)}}\big( 2M_{1}^{\left( \text{d} \right)}+1 \big)$, the beamwidth of the main lobe on the $y$-axis for PNA is upper bounded by $\frac{8}{(2M_{1}^{\left( \mathrm{d} \right)}+1)(2M_{1}^{\left( \mathrm{s} \right)}+1)}$.
When ${{M}_{y,2}}\ge 3\left( {{M}_{y,1}}+1 \right)$, the beamwidth of the main lobe on the $y$-axis for LNA is upper bounded by
$\frac{8}{\left(M_{y,1}+1\right)M_{y,2}}$.
It can be found that the beamwidth of the main lobe for UPA is $\mathcal{O}\left( \frac{1}{\sqrt{M}} \right)$, while that for PNA and LNA are $\mathcal{O}\left( \frac{1}{M} \right)$ and $\mathcal{O}\left( \frac{1}{M^{2}} \right)$, respectively.
This thus demonstrates that LNA has the highest spatial resolution, followed by PNA and finally UPA.
Therefore, when BDs are densely distributed, LNA and PNA may obtain more accurate angle estimation compared with UPA to avoid severe inter-user interference
(IUI).
\subsection{Prominent Side Lobe Height (SLH)}
%\emph{Definition 1:}For LNA, the dominating side lobes appear at ${{\Delta }_{y,m}}=\frac{2m}{{{M}_{y,1}}+1},m=\pm 1,\pm 2,\cdots ,\pm {{M}_{y,1}}$ and ${{\Delta }_{z,n}}=\frac{2n}{{{M}_{z,1}}+1},n=\pm 1,\pm 2,\cdots ,\pm {{M}_{z,1}}$,
%and the height of these lobes is $\mathrm{SLH}_\mathrm{lna}$ which is a function of $m$ and $n$.
\subsubsection{PNA}
%\emph{Definition 2:}For PNA, the dominating side lobes appear at ${{\Delta }_{y,m}}=\frac{2m}{2M_{1}^{\left( \mathrm{d} \right)}+1},m=\pm 1,\pm 2,\cdots ,\pm 2M_{1}^{\left( \mathrm{d} \right)}$,
%${{\Delta }_{z,n}}=\frac{2n}{M_{2}^{\left( \mathrm{d} \right)}},n=\pm 1,\pm 2,\cdots ,\pm \big( M_{2}^{\left( \mathrm{d} \right)}-1 \big)$,
%and these lobes have the similar height, which is $\mathrm{SLH}_\mathrm{pna}$.
For PNA, the position of the outer subarray's grating
lobes can be obtained by letting $\frac{\pi }{2}\big( 2M_{1}^{\left( \mathrm{d} \right)}+1 \big){{\Delta }_{y}}=m\pi ,m=\pm 1,\pm 2,\cdots ,\pm 2M_{1}^{\left( \mathrm{d} \right)}$,
$\frac{\pi }{2}M_{2}^{\left( \mathrm{d} \right)}{{\Delta }_{z}}=n\pi ,n=\pm 1,\pm 2,\cdots ,\pm \big( M_{2}^{\left( \mathrm{d} \right)}-1 \big)$,
and thus the $(m,n)$-th grating lobe will appear
at ${{\Delta }_{y,m}}=\frac{2m}{2M_{1}^{\left( \mathrm{d} \right)}+1},m=\pm 1,\pm 2,\cdots ,\pm 2M_{1}^{\left( \mathrm{d} \right)}$, and ${{\Delta }_{z,n}}=\frac{2n}{M_{2}^{\left( \mathrm{d} \right)}},n=\pm 1,\pm 2,\cdots ,\pm \big( M_{2}^{\left( \mathrm{d} \right)}-1 \big)$.
By taking ${{\Delta }_{y,m}}$ and  ${{\Delta }_{z,n}}$, SLH can be obtained as
\begin{equation}
\begin{split}
  \mathrm{SLH}_\mathrm{pna}& \!\approx\! \frac{1}{{{M}^{2}}}{{\Big| {{e}^{j\frac{\pi }{2}( M_{1}^{\left( \mathrm{d} \right)}+1 )( M_{2}^{\left( \mathrm{s} \right)}-1 )\frac{2n}{M_{2}^{\left( \mathrm{d} \right)}}}}M_{2}^{\left( \mathrm{s} \right)}\big( 2M_{1}^{\left( \mathrm{s} \right)}\!+\!1 \big) \Big|}^{2}} \\
 & =\frac{{{\big( M_{2}^{\left( \mathrm{s} \right)}\big( 2M_{1}^{\left( \mathrm{s} \right)}+1 \big) \big)}^{2}}}{{{M}^{2}}}.
\end{split}
\end{equation}
%Furthermore, for ${{\Delta }_{z}}=-2$, Fig. \ref{GLSNA2}, Fig. \ref{GUPA2} and Fig. \ref{GPNA2} plot the two-dimensional beam pattern of $G^{\mathrm{LSNA}}\left( {{\Delta }_{y}},-2 \right)$,  $G^{\mathrm{UPA}}\left( {{\Delta }_{y}},-2 \right)$, and $G^{\mathrm{PNA}}\left( {{\Delta }_{y}},-2 \right)$ with respect to ${\Delta }_{y}$, respectively.
%\begin{figure}[!t]
%  \centering
%  \centerline{\includegraphics[width=2.9in,height=2.2in]{GLSNA2.eps}}
%  \caption{Beam pattern of L-shaped NA with ${{\Delta }_{z}}=-2$.}
%  \label{GLSNA2}
%%  \vspace{-0.2cm}
%  \end{figure}
%\begin{figure}[!t]
%  \centering
%  \centerline{\includegraphics[width=2.9in,height=2.2in]{GUPA2.eps}}
%  \caption{Beam pattern of UPA with ${{\Delta }_{z}}=-2$.}
%  \label{GUPA2}
%%  \vspace{-0.2cm}
%  \end{figure}
%\begin{figure}[!t]
%  \centering
%  \centerline{\includegraphics[width=2.9in,height=2.2in]{GPNA2.eps}}
%  \caption{Beam pattern of PNA with ${{\Delta }_{z}}=-2$.}
%  \label{GPNA2}
%%  \vspace{-0.2cm}
%  \end{figure}
\subsubsection{LNA}
For LNA, the position of the outer subarray's grating
lobes of $y$ and $z$ can be obtained by letting $\frac{\pi }{2}\left( {{M}_{y,1}}+1 \right){{\Delta }_{y}}=m\pi ,m=\pm 1,\pm 2,\cdots ,\pm {{M}_{y,1}}$ and $\frac{\pi }{2}\left( {{M}_{z,1}}+1 \right){{\Delta }_{z}}=n\pi ,n=\pm 1,\pm 2,\cdots ,\pm {{M}_{z,1}}$, respectively, and thus the $(m,n)$-th grating lobe will appear
at ${{\Delta }_{y,m}}=\frac{2m}{{{M}_{y,1}}+1},m=\pm 1,\pm 2,\cdots ,\pm {{M}_{y,1}}$ and ${{\Delta }_{z,n}}=\frac{2n}{{{M}_{z,1}}+1},n=\pm 1,\pm 2,\cdots ,\pm {{M}_{z,1}}$.
By taking ${{\Delta }_{y,m}}$ and  ${{\Delta }_{z,n}}$, SLH can be obtained as \eqref{SLH} presented at the top of the next page.
\setcounter{TempEqCnt}{\value{equation}} % 将当前公式序号 赋给TempEqCnt
\setcounter{equation}{27} % 当前公式序号变为x，x等于长公式应有的序号减1.
\begin{figure*}[ht] %hb代表放在文章底部，%ht为放在文章顶部
\begin{equation}
{\mathrm{SLH}_\mathrm{lna}\approx \frac{1}{{{M}^{2}}}{{\left| {{\left( -1 \right)}^{m}}\left( {{M}_{y,2}}-1 \right)+{{e}^{j\left( n-m+\frac{2m}{{{M}_{y,1}}+1}-\frac{2n}{{{M}_{z,1}}+1} \right)\pi }}\left( {{\left( -1 \right)}^{n}}\left( {{M}_{z,2}}-1 \right) \right) \right|}^{2}}}.\label{SLH}
\end{equation}
\end{figure*}
When ${M}_{y,1}={M}_{z,1}$, $m=n$, ${{\mathrm{SLH}}_{\mathrm{lna}}}\approx \frac{{{\left( {{M}_{y,2}}+{{M}_{z,2}}-2 \right)}^{2}}}{{{M}^{2}}}$.

\subsubsection{Comparison of SLH for PNA and LNA}
According to the above derivation for PNA and LNA,
it can be found that the density of the grating lobe points ${{\Delta }_{y,m}}$ and ${{\Delta }_{z,n}}$ is positively correlated with the number of dense arrays $M_{1}^{\left( \mathrm{d} \right)}$, $M_{2}^{\left( \mathrm{d} \right)}$, ${M}_{y,1}$, ${M}_{z,1}$,
and the heights of prominent side lobes are positively correlated with the number of sparse arrays $M_{1}^{\left( \mathrm{s} \right)}$, $M_{2}^{\left( \mathrm{s} \right)}$, ${M}_{y,2}$, ${M}_{z,2}$.
Therefore, the number of dense and sparse arrays for NAs need to be carefully designed to balance the impact of resolution and grating lobes.
For example, when users are densely distributed, more sparse arrays can be selected for LNA and PNA to guarantee spatial resolution.
When users are sparsely distributed, fewer sparse arrays or even UPA can be deployed to reduce or eliminate the impact of grating lobes.
%Next, we further compare the beam pattern of PNA and LNA in \eqref{rhoPNA} and \eqref{rhoLS}.
%For LNA in \eqref{rhoLS}, since ${\Delta }_{y}$ and ${\Delta }_{z}$ appear independently in ${\eta }_{y}$ and ${\eta }_{z}$, and ${\rho }_{k,i}^{\mathrm{LSNA}}$ is related to the sum of ${\eta }_{y}$ and ${\eta }_{z}$, when one of ${\Delta }_{y}$ or ${\Delta }_{z}$ takes the position of the grating lobes, no matter whether the other one takes the position of the grating lobes or not, it will bring a non-negligible grating lobe value. However, for PNA in \eqref{rhoPNA}, ${\Delta }_{y}$ and ${\Delta }_{z}$ are coupled together, and ${\Delta }_{y}$ and ${\Delta }_{z}$ need to take the position of the grating lobes at the same time to bring the grating lobe value. Therefore, the grating lobes of LNA appear as a continuous line, while the grating lobes of PNA appear as a discontinuous block. Fig. \ref{GLSNA} and Fig. \ref{GPNA} are consistent with the theoretical analysis.
\section{Proposed IS$^{2}$AC-based Channel Estimation}
For SR systems with UAV swarm, obtaining CSI with conventional channel estimation for
multiple passive users introduces significant overhead,
which motivates us to propose a more efficient channel estimation method.
By exploiting the new opportunity of IS$^{2}$AC with the fact
that subspace-based super-resolution algorithms such as multiple
signal classification (MUSIC) can estimate channel parameters
accurately without requiring dedicate a priori known pilots \cite{musicDEDI},
we propose the IS$^{2}$AC-based channel estimation scheme
to use little pilots while guaranteing the communication performance.

\begin{figure}[!t]
  \centering
  \centerline{\includegraphics[width=3.1in,height=2.32in]{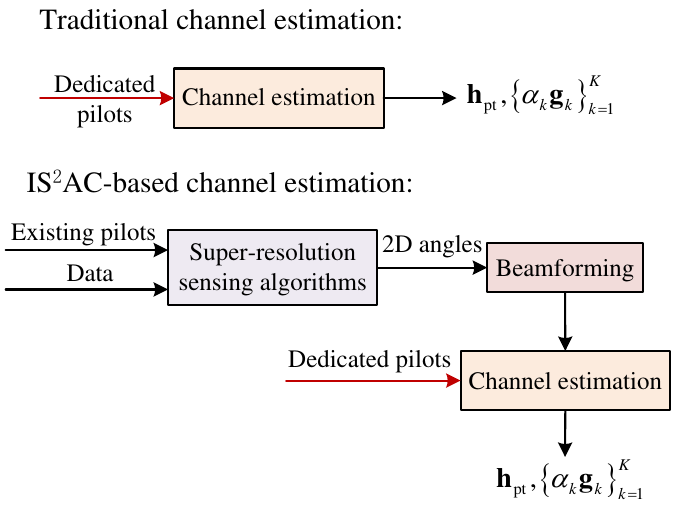}}
  \caption{The overall procedures of the proposed IS$^{2}$AC-based channel estimation and the traditional channel estimation.}\label{procedure}
  \vspace{-0.3cm}
  \end{figure}
An overview of the traditional and proposed channel estimation is illustrated in Fig. \ref{procedure}.
For the IS$^{2}$AC-based channel estimation scheme, the elevation and azimuth angles at the BS are firstly estimated by applying super-resolution sensing algorithms such as MUSIC.
Furthermore, by matching the 2D angles to obtain beamforming, channel gains are estimated accurately with very few pilots.
Since compared with the traditional channel estimation schemes without exploiting IS$^{2}$AC, there are much fewer parameters and additional beamforming gains can be achieved.
\subsection{Estimation of Elevation and Azimuth Angles}
First, we estimate the incident angles at the BS.
It is worth noting that instead of using dedicated pilots, for our proposed method, both pilot and data symbols can be used to estimate AoA associated with PT and BDs \cite{ISSACJournal}.
This is because super-resolution algorithms like MUSIC do not require the receiver to know the transmitted signal.
We denote the manifold matrix as $\mathbf{A}\left( \mathbf{\Theta },\mathbf{\Phi } \right)=\left[ \mathbf{a}\left( {{\theta }_{\mathrm{pt}}},{{\phi }_{\mathrm{pt}}} \right),\mathbf{a}\left( {{\theta }_{1}},{{\phi }_{1}} \right),\cdots ,\mathbf{a}\left( {{\theta }_{K}},{{\phi }_{K}} \right) \right]\in {{\mathbb{C}}^{M\times \left( K+1 \right)}}$ for both NA deployments.
\subsubsection{LNA}
The process of angle estimation for LNA is firstly introduced.
Denote the pilot sequence sent by the PT as ${{\psi }_{\mathrm{pt}}}\left( n \right)$ satisfying $\sum\limits_{n=1}^{\tau }{{{\left| {{\psi }_{\mathrm{pt}}}\left( n \right) \right|}^{2}}}=\tau$,
and the pilot sequence sent by the BD as ${{\psi }_{k}}\left( n \right)$ satisfying $\sum\limits_{n=1}^{\tau }{{{\left| {{\psi }_{k}}\left( n \right) \right|}^{2}}}=\tau$ for $k=1,...,K$.
For BS, the observed signals at NA of the $z$ axis during the pilot and data symbol durations are respectively written as
\begin{equation}
{{{\mathbf{y}}_{z,t}}\left( n \right)={{\mathbf{A}}_{z}}\left( \mathbf{\Theta } \right){{\bm{\psi}}_{\mathrm{com}}}(n)+\mathbf{u}_{z,t}\left( n \right),\label{yzpilot}}
\end{equation}
\begin{equation}
{{{\mathbf{y}}_{z,d}}\left( n \right)={{\mathbf{A}}_{z}}\left( \mathbf{\Theta } \right){{\mathbf{s}}_{\mathrm{com}}}(n)+\mathbf{u}_{z,d}\left( n \right),\label{yz}}
\end{equation}
where ${{\mathbf{A}}_{z}}\left( \mathbf{\Theta } \right)=\left[ {{\mathbf{a}}_{z}}\left( {{\theta }_{\mathrm{pt}}} \right),{{\mathbf{a}}_{z}}\left( {{\theta }_{1}} \right),\cdots ,{{\mathbf{a}}_{z}}\left( {{\theta }_{K}} \right) \right]\in {{\mathbb{C}}^{M_z\times \left( K+1 \right)}}$ is the manifold matrix of NA at the $z$-axis,
${{\bm{\psi }}_{\mathrm{com}}}(n)=\sqrt{{{P}_{d}}}[{{\gamma }_{\mathrm{pt}}}{{\psi }_{\mathrm{pt}}}\left( n \right),{{\alpha }_{1}}{{\beta }_{1}}{{\psi }_{1}}\left( n \right){{\psi }_{\mathrm{pt}}}\left( n \right)
,\cdots ,\\
{{\alpha }_{K}}{{\beta }_{K}}{{\psi }_{K}}\left( n \right){{\psi }_{\mathrm{pt}}}\left( n \right)]^{\mathrm{T}}
\in {{\mathbb{C}}^{\left( K+1 \right)\times 1}}$,
${{\mathbf{s}}_{\mathrm{com}}}(n)=\sqrt{{{P}_{d}}}[ {{\gamma }_{\mathrm{pt}}}{s}\left( n \right),{{\alpha }_{1}}{{\beta }_{1}}{{c}_{1}}\left( n \right){s}\left( n \right),\cdots,
{{\alpha }_{K}}{{\beta }_{K}}{{c}_{K}}\left( n \right){s}\left( n \right) ]^{\mathrm{T}}\in {{\mathbb{C}}^{\left( K+1 \right)\times 1}}$,
$\mathbf{u}_{z,t}(n)\in \mathbb{C}^{M_z\times 1}$ and $\mathbf{u}_{z,d}(n)\in \mathbb{C}^{M_z\times 1}$ denote the CSCG noise with zero mean and power ${{\sigma }^{2}}$.
Given the non-uniform spatial sampling of the NA,
classic AoA estimation cannot be directly performed.
However, by exploiting the potential of NA,
the virtual receive signal model can be used here to tackle this problem \cite{NA1}.
The process of angle estimation is described as follows.
Firstly, calculate the covariance matrix of the observed signal in \eqref{yzpilot} and \eqref{yz} as
\begin{equation}
{\mathbf{R}_{z}=\mathbb{E}\left[ \mathbf{y}_{z}\left( n \right){{\mathbf{y}}_{z}^{\mathrm{H}}}\left( n \right) \right]=\mathbf{A}_{z}\left( \mathbf{\Theta } \right){{\mathbf{R}}_{s}}{{\mathbf{A}}_{z}^{\mathrm{H}}\left( \mathbf{\Theta } \right)}+{{\sigma }^{2}}{{\mathbf{I}}_{M_z}}},
\end{equation}
where $\mathbf{y}_{z}\left( n \right)$ is taken from $\mathbf{y}_{z,t}\left( n \right)$ and $\mathbf{y}_{z,d}\left( n \right)$, ${{\mathbf{R}}_{s}}={{P}_{d}}\mathrm{diag}\big\{ {{\left| {{\gamma }_{\mathrm{pt}}} \right|}^{2}},{{\left| {{\alpha }_{1}}{{\beta }_{1}} \right|}^{2}},\cdots ,{{\left| {{\alpha }_{K}}{{\beta }_{K}} \right|}^{2}} \big\}\in {{\mathbb{C}}^{\left( K+1 \right)}}$ is the covariance matrix of ${{\bm{\psi }}_{\mathrm{com}}}(n)$ and ${{\mathbf{s}}_{\mathrm{com}}(n)}$.
Note that in practice, the expectation $\mathbb{E}\left[ \mathbf{y}_{z}\left( n \right){{\mathbf{y}}_{z}^{\mathrm{H}}}\left( n \right) \right]$
is usually performed via the sample average $\frac{1}{N}\sum\limits_{n=1}^{N}{{{\mathbf{y}}_{z}}\left( n \right)\mathbf{y}_{z}^{\mathrm{H}}\left( n \right)}$,
where $N$ is the number of samples with large value.

Then, $\mathbf{R}_{z}$ is vectorized to obtain
\begin{equation}
{\mathbf{r}_{z}=\mathrm{vec}\left( \mathbf{R}_{z} \right)=\left( {{\mathbf{A}^{*}_{z}\left( \mathbf{\Theta } \right)}}\odot \mathbf{A}_{z}\left( \mathbf{\Theta } \right) \right){{\mathbf{r}}_{s}}+{{\sigma }^{2}}\overset{\to }{\mathop{\mathbf{1}_z}}\label{r},}
\end{equation}
where ${{\mathbf{r}}_{s}}={{P}_{d}}{\big[ {{\left| {{\gamma }_{\mathrm{pt}}} \right|}^{2}},{{\left| {{\alpha }_{1}}{{\beta }_{1}} \right|}^{2}},\cdots ,{{\left| {{\alpha }_{K}}{{\beta }_{K}} \right|}^{2}} \big]}^{\mathrm{T}}\in {{\mathbb{C}}^{\left( K+1 \right)\times 1}}$ can be regarded as the equivalent receive signal, and ${{\sigma }^{2}}\overset{\to }{\mathop{\mathbf{1}_z}}\,={{\sigma }^{2}}{{\left[ \mathbf{e}_{1}^{\mathrm{T}},\cdots ,\mathbf{e}_{M_z}^{\mathrm{T}} \right]}^{\mathrm{T}}}$ is the equivalent receive noise where $\mathbf{e}_{i}$ is a column vector of all zeros except the $i$-th element being one.
The conjugate and KR product $\odot$ operations act as the difference of the positions of the physical antennas, thus forming
a virtual array whose antennas locate at ${\mathcal{D}_{\mathrm{vir},z}}=\left\{ {{d}_{z,i}}-{{d}_{z,j}},i,j=1,2,\cdots ,M_z \right\}$.
Furthermore, by removing repeated rows and sorting the virtual array antennas,
an equivalent steering matrix can be obtained, with the receiving antennas located at distinct values of ${\mathcal{D}_{r_z}}$.
It has been proved that when $M_z$ is even and ${{M}_{z,1}}={{M}_{z,2}}=\frac{M_z}{2}$,
by using only $M_z$ physical antennas, ${\mathcal{D}_{r_z}}$ can equivalently form a compact ULA with at most $\frac{{{M}_{z}^{2}}-2}{2}+M_{z}$
virtual antennas \cite{NA1}.
Accordingly, the DoFs and array aperture are dramatically enhanced, which can improve the sensing performance.
Since \eqref{r} is a single snapshot signal,
spatial smoothing need to be used to recover the rank of the covariance
matrix of $\mathbf{r}_{z}$ \cite{NA1}.
Then, classic super-resolution algorithms such as MUSIC \cite{musicDEDI} can be performed to estimate the elevation angles of the PT and $K$ BDs,
which can be respectively estimated as ${{\hat{\theta }}_{\mathrm{pt}}}$ and ${{\hat{\theta }}_{k}}, k=1,...,K$.
%Denote the MUSIC spectrum as ${P}_{\mathrm{MUSIC}}\left( {\theta } \right)$.
%Based on the spectral peaks estimated by MUSIC in descending order,
%reordering the BDs as $1,2,...,K$, which satisfies ${{P}_{\mathrm{MUSIC}}}\left( {{\theta }_{1}} \right)\ge {{P}_{\mathrm{MUSIC}}}\left( {{\theta }_{2}} \right)\ge \cdots \ge {{P}_{\mathrm{MUSIC}}}\left( {{\theta }_{K}} \right)$.
%Due to the double-hop signal attenuation and backscattering loss involved, the backscattering link of the
%BD is usually very weak compared to the direct link, which results ${{P}_{\mathrm{MUSIC}}}\left( {{\theta }_{\mathrm{ue}}} \right)\ge {{P}_{\mathrm{MUSIC}}}\left( {{\theta }_{1}} \right)$.
Further, we can perform similar operations as above to estimate the incident angles ${\hat{\varphi }_{\mathrm{pt}}}$, ${\hat{\varphi }_{k}}, k=1,...,K$ by using the received signal along the $y$ axis.
%Assume all channels follow block-fading law, which are constant for time interval $T$, after which they change to independent
%values that holds for another interval, and so on.
%For one time interval $T$, let $\tau$ denote the length of uplink training sequence with $1\le \tau \le K+1$,
%then the length of data transmission sequence is $T-\tau$.
However, since the two sets of angles $\big\{ {\hat{\theta }_{\mathrm{pt}}}, {\hat{\theta }_{k}}, k=1,...,K \big\}$ and $\big\{ {\hat{\varphi }_{\mathrm{pt}}}, {\hat{\varphi }_{k}}, k=1,...,K \big\}$ are separately estimated, they need to be pair-matched in order to be utilized for the identification of 2-D directions.
In order to solve this issue, an extra procedure is used here to
implement the pairing of 1-D AOA estimates in \cite{Lshaped1}.
The steps are briefly described as follows:
Firstly, substitute the estimated $\big\{ {\hat{\theta }_{\mathrm{pt}}}, {\hat{\theta }_{k}}, k=1,...,K \big\}$ into \eqref{r} that yields $\hat{\mathbf{r}}_{z}$.
Then, remove the repeated rows from $\hat{\mathbf{r}}_{z}$ and sort
them as above to yield
\begin{equation}
{\mathbf{\bar{r}}_{z}=\left( \mathbf{\bar{A}}_{z}^{*}\left( \mathbf{\Theta } \right)\odot {{{\mathbf{\bar{A}}}}_{z}}\left( \mathbf{\Theta } \right) \right){{\mathbf{r}}_{s}}+{{\sigma }^{2}}{\mathbf{{e}}}_{{{M}_{z,2}}\left( {{M}_{z,1}}+1 \right)}},\label{barr}
\end{equation}
where ${\mathbf{{e}}}_{{{M}_{z,2}}\left( {{M}_{z,1}}+1 \right)}\in {{\mathbb{C}}^{2{{M}_{z,2}}\left( {{M}_{z,1}}+1 \right)-1}}$ denotes the column vector of all zeros except one 1 at the ${{M}_{z,2}}\left( {{M}_{z,1}}+1 \right)$-th position.
Then, eliminate the ${{M}_{z,2}}\left( {{M}_{z,1}}+1 \right)$-th
element of $\mathbf{\bar{r}}_{z}$ to remove the effect of noise, \eqref{barr} becomes $\mathbf{\tilde{r}}_{z}=\big( \mathbf{\tilde{A}}_{z}^{*}\left( \mathbf{\Theta } \right)\odot {{{\mathbf{\tilde{A}}}}_{z}}\left( \mathbf{\Theta } \right) \big){{\mathbf{r}}_{s}}$, where ${{\mathbf{\tilde{A}}}_{z}}\left( \mathbf{\Theta } \right)$ is constructed by removing the
${{M}_{z,2}}\left( {{M}_{z,1}}+1 \right)$-th row from ${{\mathbf{\bar{A}}}_{z}}\left( \mathbf{\Theta } \right)$.
The vector ${{\mathbf{r}}_{s}}$ can be estimated by
\begin{equation}
\setlength\abovedisplayskip{1pt}
\setlength\belowdisplayskip{1pt}
{{{\mathbf{\hat{r}}}_{s}}={{\big( \mathbf{\tilde{A}}_{z}^{*}\left( \mathbf{\Theta } \right)\odot {{{\mathbf{\tilde{A}}}}_{z}}\left( \mathbf{\Theta } \right) \big)}^{\dagger }}\mathbf{\tilde{r}}_{z},}
\end{equation}
where ${{\left( \cdot \right)}^{\dagger }}$ stands for the pseudoinverse operator.
Then, the covariance matrix of ${{\bm{\psi }}_{\mathrm{com}}}(n)$ and ${{\mathbf{s}}_{\mathrm{com}}(n)}$ can be estimated as ${{\mathbf{\hat{R}}}_{s}}=\mathrm{diag}\left( {{{\mathbf{\hat{r}}}}_{s}} \right)$.
Furthermore, a permutation matrix $\mathbf{T}\in {{\mathbb{R}}^{\left( K+1 \right)}}$ is introduced to match one dimensional (1D) angles,
where ${{\left[ \mathbf{T} \right]}_{i,j}}\in \left\{ 0,1 \right\}$, $\sum\limits_{i=1}^{K+1}{{{\left[ \mathbf{T} \right]}_{i,j}}}=1$ and $\sum\limits_{j=1}^{K+1}{{{\left[ \mathbf{T} \right]}_{i,j}}}=1$.
By using the permutation matrix $\mathbf{T}$,
the incident angles $\big\{ {\hat{\varphi }_{\mathrm{pt}}}, {\hat{\varphi }_{k}}, k=1,...,K \big\}$
can be rearranged corresponding to $\big\{ {\hat{\theta }_{\mathrm{pt}}}, {\hat{\theta }_{k}}, k=1,...,K \big\}$, i.e.,
$\big[ {{{\hat{\varphi }}}_{\mathrm{pt}}^{\prime }},{{{\hat{\varphi }}}_{1}^{\prime }},\cdots ,{{{\hat{\varphi }}}_{K}^{\prime }} \big]=\big[ {{{\hat{\varphi }}}_{\mathrm{pt}}},{{{\hat{\varphi }}}_{1}},\cdots ,{{{\hat{\varphi }}}_{K}} \big]\mathbf{T}$.
Denote the manifold matrix of NA at the $y$-axis as ${{\mathbf{A}}_{y}}\left( \mathbf{\Phi } \right)=\left[ {{\mathbf{a}}_{z}}\left( {{\varphi }_{\mathrm{pt}}} \right),{{\mathbf{a}}_{z}}\left( {{\varphi }_{1}} \right),\cdots ,{{\mathbf{a}}_{z}}\left( {{\varphi }_{K}} \right) \right]\in {{\mathbb{C}}^{M_y\times \left( K+1 \right)}}$.
Substituting $\big\{ {\hat{\varphi }_{\mathrm{pt}}^{\prime }}, {\hat{\varphi }_{k}^{\prime }}, k=1,...,K \big\}$ into ${{{\mathbf{{A}}}}_{y}}\left( \mathbf{\Phi } \right)$, we have
\begin{equation}
{{{\mathbf{A}}_{y}^{\prime }}\big( {\mathbf{\hat{\Phi }}} \big)={{\mathbf{A}}_{y}}\big( {\mathbf{\hat{\Phi }}} \big)\mathbf{T}\approx {{\mathbf{A}}_{y}}\left( {\mathbf{\Phi }} \right)}.
\end{equation}
The cross-covariance matrix between the observed signals at
NA of $z$ axis $\mathbf{y}_{z}\left( n \right)$ and the observed signals at NA of $y$ axis $\mathbf{y}_{y}\left( n \right)$ is written as
\begin{equation}
\begin{split}
\setlength\abovedisplayskip{1pt}
\setlength\belowdisplayskip{1pt}
  {{\mathbf{R}}_{yz}}& ={{\mathbb{E}}_{yz}}\left[ {{\mathbf{y}}_{y}}\left( n \right)\mathbf{y}_{z}^{\mathrm{H}}\left( n \right) \right]={{\mathbf{A}}_{y}}\left( \mathbf{\Phi } \right){{\mathbf{R}}_{s}}\mathbf{A}_{z}^{\mathrm{H}}\left( \mathbf{\Theta } \right) \\
 & \approx {{\mathbf{A}}_{y}}\big( {\mathbf{\hat{\Phi }}} \big)\mathbf{T}{{{\mathbf{\hat{R}}}}_{s}}\mathbf{A}_{z}^{\mathrm{H}}\big( {\mathbf{\hat{\Theta }}} \big).
\end{split}
\end{equation}
The permutation matrix $\mathbf{T}$ can be obtained by solving the following
minimization problem:
\begin{equation}
{\underset{\mathbf{T}}{\mathop{\min }}\,{{\left\| {{\mathbf{R}}_{yz}}-{{\mathbf{A}}_{y}}\big( {\mathbf{\hat{\Phi }}} \big)\mathbf{T}{{{\mathbf{\hat{R}}}}_{s}}\mathbf{A}_{z}^{\mathrm{H}}\big( {\mathbf{\hat{\Theta }}} \big) \right\|}_{F}}}.
\end{equation}
Then the paired angles $\big\{ {\hat{\varphi }_{\mathrm{pt}}}, {\hat{\varphi }_{k}}, k=1,...,K \big\}$ can be obtained.

With the paired 2D estimated angles $\big\{ {\hat{\theta }_{\mathrm{pt}}}, {\hat{\varphi }_{\mathrm{pt}}}\big\}$ and $\big\{{{\hat{\theta }}_{k}},{{\hat{\varphi }}_{k}}\big\}_{k=1}^{K}$,
the azimuth angles of PT and BDs can be respectively calculated by
\begin{equation}
\setlength\abovedisplayskip{1pt}
\setlength\belowdisplayskip{1pt}
{{{\hat{\phi} }_{\mathrm{pt}}}=\arccos \Big( \frac{\cos {{\hat{\varphi} }_{\mathrm{pt}}}}{\cos {{\hat{\theta} }_{\mathrm{pt}}}} \Big)},
\end{equation}
\begin{equation}
\setlength\abovedisplayskip{1pt}
\setlength\belowdisplayskip{1pt}
{{{\hat{\phi} }_{k}}=\arccos \Big( \frac{\cos {{\hat{\varphi} }_{k}}}{\cos {{\hat{\theta} }_{k}}} \Big)},k=1,...,K.
\end{equation}
\subsubsection{PNA}
For PNA, the observed signals at the BS during the pilot and data symbol durations can be respectively written as
\begin{equation}
\setlength\abovedisplayskip{1pt}
\setlength\belowdisplayskip{1pt}
{{{\mathbf{y}}_{t}}\left( n \right)={{\mathbf{A}}}\left( \mathbf{\Theta },\mathbf{\Phi } \right){{\bm{\psi}}_{\mathrm{com}}}(n)+\mathbf{u}_{t}\left( n \right),\label{yPNApilot}}
\end{equation}
\begin{equation}
\setlength\abovedisplayskip{1pt}
\setlength\belowdisplayskip{1pt}
{{{\mathbf{y}}_{d}}\left( n \right)={{\mathbf{A}}}\left( \mathbf{\Theta },\mathbf{\Phi } \right){{\mathbf{s}}_{\mathrm{com}}}(n)+\mathbf{u}_{d}\left( n \right),\label{yPNA}}
\end{equation}
where $\mathbf{A}\left( \mathbf{\Theta },\mathbf{\Phi } \right)$,
${{\bm{\psi}}_{\mathrm{com}}}(n)$ and ${{\mathbf{s}}_{\mathrm{com}}}(n)$ are defined in Section III A above,
$\mathbf{u}_{t}(n)\in \mathbb{C}^{M\times 1}$ and $\mathbf{u}_{d}(n)\in \mathbb{C}^{M\times 1}$ denote the CSCG noise with zero mean and power ${{\sigma }^{2}}$.
Note that \eqref{yPNApilot} and \eqref{yPNA} have similar forms to \eqref{yzpilot} and \eqref{yz} but correspond to different array response vectors with different array configurations.

Then the covariance matrix of \eqref{yPNApilot} and \eqref{yPNA} can be written as
\begin{equation}
\begin{split}
\setlength\abovedisplayskip{1pt}
\setlength\belowdisplayskip{1pt}
  \mathbf{R}& =\mathbb{E}\left[ {{\mathbf{y}}}\left( n \right)\mathbf{y}^{\mathrm{H}}\left( n \right) \right] \\
 & =\mathbf{A}\left( \mathbf{\Theta },\mathbf{\Phi } \right){{\mathbf{R}}_{s}}{{\mathbf{A}}^{\mathrm{H}}}\left( \mathbf{\Theta },\mathbf{\Phi } \right)+{{\sigma }^{2}}{{\mathbf{I}}_{M}},
\end{split}
\end{equation}
where $\mathbf{y}\left( n \right)$ is taken from $\mathbf{y}_{t}\left( n \right)$ and $\mathbf{y}_{d}\left( n \right)$,
${\mathbf{R}}_{s}$ is the covariance matrix of ${{\bm{\psi}}_{\mathrm{com}}}(n)$ and ${{\mathbf{s}}_{\mathrm{com}}}(n)$.
Then, vectorize $\mathbf{R}$ to obtain
\begin{equation}
\setlength\abovedisplayskip{1pt}
\setlength\belowdisplayskip{1pt}
{\mathbf{r}=\mathrm{vec}\left( \mathbf{R} \right)=\left( {{\mathbf{A}}^{*}}\left( \mathbf{\Theta },\mathbf{\Phi } \right)\odot \mathbf{A}\left( \mathbf{\Theta },\mathbf{\Phi } \right) \right){{\mathbf{r}}_{s}}+{{\sigma }^{2}}\overset{\to }{\mathop{\mathbf{1}}}\,,}
\end{equation}
where ${{\mathbf{r}}_{s}}$ is the equivalent receive signal and noise defined in Section III A 1),
${{\sigma }^{2}}\overset{\to }{\mathop{\mathbf{1}}}\,={{\sigma }^{2}}{{\left[ \mathbf{e}_{1}^{\mathrm{T}},\cdots ,\mathbf{e}_{M}^{\mathrm{T}} \right]}^{\mathrm{T}}}$ is the equivalent receive noise.
%The conjugate and KR product $\odot$ operations act as the difference of the positions of the physical
%antennas, thus forming a virtual array whose antennas locate
%at $\mathcal{D}_{\mathrm{vir,PNA}}$ defined in Section II B.
Furthermore, by selecting and sorting the virtual array antennas,
an equivalent steering matrix can be obtained, with the receiving antennas located at distinct values of ${\mathcal{D}_{r}}$.
To recover the rank of the covariance matrix,
2D spatial smoothing is performed \cite{NA2D2}.
Then, classic super-resolution algorithms such as MUSIC \cite{musicDEDI}
is used to estimate the 2D elevation and azimuth angles of the
UE and $K$ BDs, which can be respectively estimated as ${{\hat{\theta }}_{\mathrm{pt}}}$, ${{\hat{\phi }}_{\mathrm{pt}}}$, ${{\hat{\theta }}_{k}}$, and ${{\hat{\phi }}_{k}}, k=1,...,K$.
\subsection{Estimation of Channel Gains}
To obatin the CSI of the direct and backscatter links,
we still need to estimate the complex-valued path coefficients of the PT and BDs.
Assume that the pilot symbols backscattered by the $k$-th BD are equal to 1,
while the pilot sequence ${\psi}(n)$ are sent by the PT with $\sum\limits_{i=1}^{{{\tau }}}{{{\left| \psi \left( i \right) \right|}^{2}}}={{\tau }} $, which is known to the receiver.
We construct the receive beamforming matrix,
which consists of the beamforming vectors that match
the PT and $K$ BDs, as
\begin{small}
\begin{equation}
\begin{split}
\setlength\abovedisplayskip{1pt}
\setlength\belowdisplayskip{1pt}
\mathbf{W}& ={{\bigg[ \frac{\mathbf{a}\big( {{{\hat{\theta }}}_{\mathrm{pt}}},{{{\hat{\phi }}}_{\mathrm{pt}}} \big)}{\big\| \mathbf{a}\big( {{{\hat{\theta }}}_{\mathrm{pt}}},{{{\hat{\phi }}}_{\mathrm{pt}}}\big) \big\|},\frac{\mathbf{a}\big( {{{\hat{\theta }}}_{1}},{{{\hat{\phi }}}_{1}} \big)}{\big\| \mathbf{a}\big( {{{\hat{\theta }}}_{1}},{{{\hat{\phi }}}_{1}} \big) \big\|}\cdots ,\frac{\mathbf{a}\big( {{{\hat{\theta }}}_{K}},{{{\hat{\phi }}}_{K}} \big)}{\big\| \mathbf{a}\big( {{{\hat{\theta }}}_{K}},{{{\hat{\phi }}}_{K}} \big) \big\|} \bigg]}^{\mathrm{H}}}\\
& =\frac{1}{\sqrt{M}}{{\mathbf{A}}^{\mathrm{H}}}\big( \hat{\bm{\Theta }},\hat{\bm{\Phi }} \big)\in {{\mathbb{C}}^{\left( K+1 \right)\times M}}.
\end{split}
\end{equation}
\end{small}
Let $P_t$ denote the transmit power by the PT during the pilot transmission phase.
Further denote $\bm{\gamma }={{\left[ {{\gamma }_{\mathrm{pt}}},{{\alpha }_{1}}{{\beta }_{1}},\cdots ,{{\alpha }_{K}}{{\beta }_{K}} \right]}^{\mathrm{T}}}\in {{\mathbb{C}}^{\left( K+1 \right)\times 1}}$,
then the resulting signal at the BS can be written as
\begin{small}
\begin{equation}
\begin{split}
  & \mathbf{y}_{t}\left( n \right)=\sqrt{{P}_{t}}\mathbf{W}{\mathbf{A}}\left( {\bm{\Theta }},{\bm{\Phi }} \right)\bm{\gamma }\psi \left( n \right)+\mathbf{W}\mathbf{u}_{t}\left( n \right) \\
 & =\frac{\sqrt{{{P}_{t}}}}{\sqrt{M}}{{\mathbf{A}}^{\mathrm{H}}}\big( \hat{\bm{\Theta }},\hat{\bm{\Phi }} \big)\mathbf{A}\big( \bm{\Theta },\bm{\Phi } \big)\bm{\gamma }\psi \left( n \right)+\frac{1}{\sqrt{M}}{{\mathbf{A}}^{\mathrm{H}}}\big( \hat{\bm{\Theta }},\hat{\bm{\Phi }} \big)\mathbf{u}_{t}\left( n \right),
\end{split}
\end{equation}
\end{small}where $\mathbf{u}_{t}(n)$ denotes the CSCG noise with zero-mean and variance ${\sigma }^{2}$.
Since the pilot symbol ${\psi}(n)$ is known, $\mathbf{y}_{t}(n)$ can be projected to ${\psi}^{*}(n)$, which results in
\begin{small}
\begin{equation}
\begin{split}
\setlength\abovedisplayskip{0.5pt}
\setlength\belowdisplayskip{0.5pt}
  \mathbf{y}_{t}^{\prime }& =\frac{1}{\tau }\sum\limits_{i=1}^{\tau }{\mathbf{y}_{t}\left( i \right)}{{\psi }^{*}}\left( i \right) \\
 & \!=\!\frac{\sqrt{{{P}_{t}}}}{\sqrt{M}}{{\mathbf{A}}^{\mathrm{H}}}\big( {\mathbf{\hat{\Theta }}},\hat{\bm{\Phi }} \big)\mathbf{A}\big( \mathbf{\Theta },\bm{\Phi } \big)\bm{\gamma }\!+\!\frac{1}{\sqrt{M}\tau}{{\mathbf{A}}^{\mathrm{H}}}\big( {\bm{\hat{\Theta }}},\hat{\bm{\Phi }} \big)\mathbf{u}_{t}^{\prime },
\end{split}
\end{equation}
\end{small}where $\mathbf{u}_{t}^{\prime }=\sum\limits_{i=1}^{\tau }{{\mathbf{u}_{t}}}\left( i \right){{\psi }^{*}}\left( i \right)$ is the resulting noise vector, with
${\mathbf{u}_{t}^{\prime }}\sim\mathcal{C}\mathcal{N}\left( 0, {\tau }{{\sigma }^{2}}\mathbf{I}_{M}\right)$.
%If the angle is estimated without error, i.e., ${\mathbf{\hat{\Theta }}}=\mathbf{\Theta }$, ${\mathbf{\hat{\Phi }}}=\mathbf{\Phi }$,
The least squares (LS) estimation of $\bm{\gamma }$ can be written as
\begin{equation}
\setlength\abovedisplayskip{1pt}
\setlength\belowdisplayskip{1pt}
{\bm{\hat{\gamma }}=\frac{\sqrt{M}}{\sqrt{{P}_{t}}}{{\left( {{\mathbf{A}}^{\mathrm{H}}}\big( {\mathbf{\hat{\Theta }}},{\bm{\hat{\Phi }}} \big)\mathbf{A}\big( {\mathbf{\hat{\Theta }}},\bm{\hat{\Phi} } \big) \right)}^{-1}}\mathbf{y}_{t}^{\prime }}.
\end{equation}

After the incident angles and channel gains are obtained, the direct link and cascaded channels can be respectively estimated as
${{\mathbf{\hat{h}}}_{\mathrm{pt}}}={{\hat{\gamma }}_{\mathrm{pt}}}\mathbf{a}\big( {{{\hat{\theta }}}_{\mathrm{pt}}},{{{\hat{\phi }}}_{\mathrm{pt}}} \big)$,
${{\hat{\alpha }}_{k}}{{\mathbf{\hat{g}}}_{k}}={{\hat{\alpha }}_{k}}{{\hat{\beta }}_{k}}\mathbf{a}\big( {{{\hat{\theta }}}_{k}},{{{\hat{\phi }}}_{k}} \big),k=1,...,K$.

In order to further calculate the channel estimation performance of UE and UAVs for the proposed channel estimation method,
we define the whole channel as ${{\mathbf{h}}_{\mathrm{sum}}}=\mathbf{A}(\mathbf{\Theta },\mathbf{\Phi} )\bm{\gamma }$.
If the angles are estimated perfectly, i.e., ${\mathbf{\hat{\Theta }}}=\mathbf{\Theta }$, ${\mathbf{\hat{\Phi }}}=\mathbf{\Phi }$, the estimation error of ${{\mathbf{h}}_{\mathrm{sum}}}$ is written as
${{\mathbf{\tilde{h}}}_{\mathrm{sum}}}=\mathbf{A}(\mathbf{\hat{\Theta }},\mathbf{\hat{\Phi }})\bm{\hat{\gamma }}-\mathbf{A}(\mathbf{\hat{\Theta }},\mathbf{\hat{\Phi }})\bm{\gamma }=\mathbf{A}(\mathbf{\hat{\Theta }},\mathbf{\hat{\Phi }})\bm{\tilde{\gamma }}$.
Following a similar derivation as \cite{wcnc}, the MSE of ${{\mathbf{h}}_{\mathrm{sum}}}$ can be written as
\begin{equation}
\setlength\abovedisplayskip{1pt}
\setlength\belowdisplayskip{1pt}
{\mathbb{E}\big[ {{\big\| {{{\mathbf{\tilde{h}}}}_{\mathrm{sum}}} \big\|}^{2}} \big]=\frac{\left( K+1 \right){{\sigma }^{2}}}{\tau {{P}_{t}}}}.\label{MSEhsum}
\end{equation}
Similar to the analysis in \cite{wcnc}, the MSE of ${{\mathbf{\tilde{h}}}_{\mathrm{sum}}}$ for conventional channel estimation method based on channel training is
$\frac{M{{\sigma }^{2}}}{\tau {{P}_{t}}}$.
When the number of PT and BDs $K+1$ is larger than the number of antennas $M$,
our proposed IS$^{2}$AC-based channel estimation method can achieve better estimation performance compared with the traditional scheme.
Note that compared with our previous work on the IS$^{2}$AC-based channel estimation scheme \cite{wcnc,ISSACJournal}, this work focuses on the scenario where the sparse array is deployed at the BS, so as to obtain more accurate angle estimation with densely located BDs.
\section{Simulation Results}
In this section, simulation results are provided to compare the performance of PNA, LNA and UPA for the IS$^{2}$AC-SR
system with the low-altitude UAV swarm.
We assume that the incident angles of the BS from $K$ BDs are
uniformly distributed on $[-\theta_\mathrm{max}, \theta_\mathrm{max}]$ and $[-\varphi_\mathrm{max}, \varphi_\mathrm{max}]$, respectively.
The path gains $\gamma_{\mathrm{pt}}$, $\alpha_{k}$ and $\beta_{k}$ are respectively generated as following:
$\gamma_{\mathrm{pt}}\sim\mathcal{C}\mathcal{N}\left( 0, 1\right)$, $\alpha_{k}\beta_{k}\sim\mathcal{C}\mathcal{N}\left( 0, 0.1\right),k=1,...,K$.
Define ${{\bar{P}}_{t}}=\frac{{{P}_{t}}}{{{\sigma }^{2}}}$ and ${{\bar{P}}_{d}}=\frac{{{P}_{d}}}{{{\sigma }^{2}}}$ as the transmit signal-to-noise ratio (SNR) by the PT in pilot and data transmission phases,
respectively. Unless otherwise stated, we set the number of the BDs as $K=3$, ${{\bar{P}}_{t}}={{\bar{P}}_{d}}= 20$ dB, ${{\theta }_{\mathrm{max}}}={{\varphi }_{\mathrm{max}}}=10^\circ$, and the number of antennas as $M=32$.

\begin{figure}[!t]
  \centering
  \centerline{\includegraphics[width=2.3in,height=1.7in]{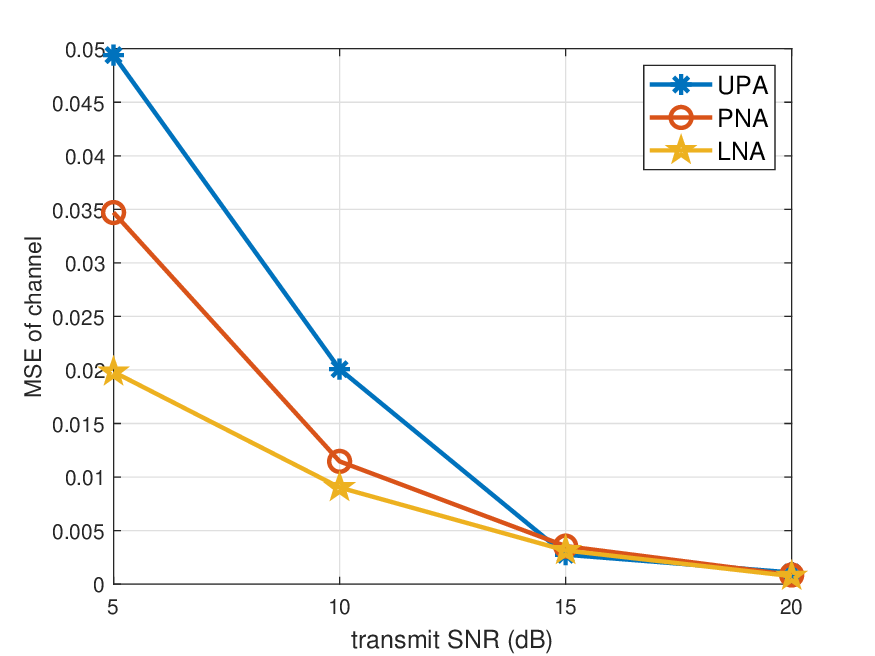}}
  \caption{MSE versus transmit SNR ${\bar{P}}_{t}$.}
  \label{MMSEvsPt}
  \vspace{-0.2cm}
  \end{figure}

\begin{figure}[!t]
  \centering
  \centerline{\includegraphics[width=2.3in,height=1.7in]{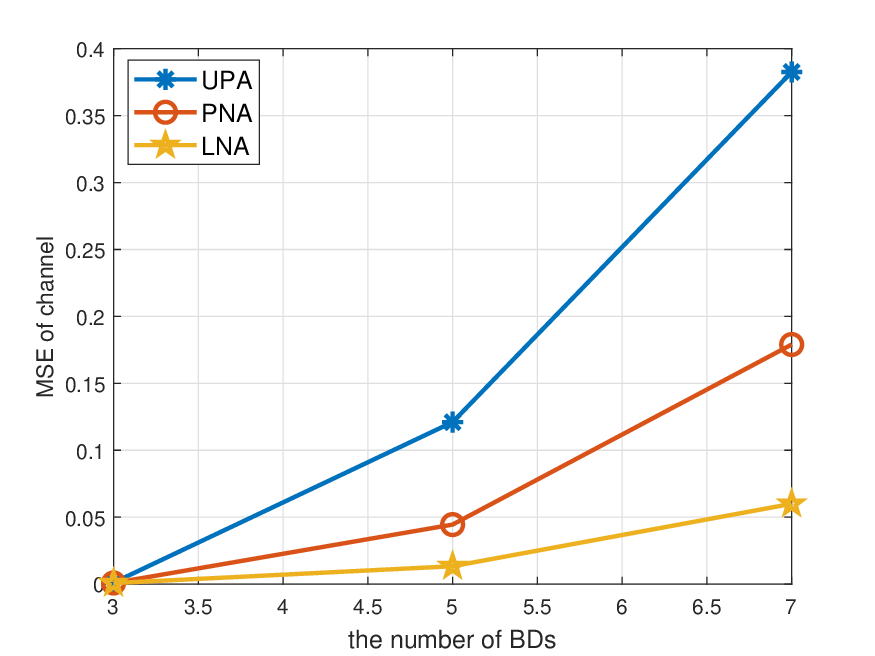}}
  \caption{MSE versus the number of BDs $K$.}
  \label{MMSEvsK}
  \vspace{-0.2cm}
  \end{figure}

Define the sum MSE of channel gains as ${{\mathrm{MSE}}_{\gamma }}=\frac{1}{S}\sum\limits_{s=1}^{S}{{{\left( {{{\bm{\hat{\gamma }}}}_{s}}-\bm{\gamma } \right)}^{2}}}$,
where $S$ is the number of Monte-Carlo simulations.
Fig. \ref{MMSEvsPt}, and \ref{MMSEvsK} plot ${{\mathrm{MSE}}_{\gamma}}$
for PNA, LNA and UPA versus transmit SNR ${{\bar{P}}_{t}}$ and the number of BDs $K$, respectively.
%It can be observed from Fig. \ref{MMSEvsM} that when the number of antennas is limited, i.e., 16 elements, the UPA exhibits significantly inferior estimation performance compared to LNA and PNA.
%In addition, LNA demonstrates a slight performance advantage over PNA, which is attributed to its capability of achieving a larger array aperture.
%However, when the number of antennas exceeds a threshold, i.e., $>$32 elements, all three array configurations achieve accurate channel estimation performance, owing to the sufficient spatial degrees of freedom provided by the large-scale array aperture.
It is observed from Fig. \ref{MMSEvsPt} that when the transmit SNR is low, LNA achieves better estimation performance than PNA, which in turn outperforms the UPA.
It is also observed from Fig. \ref{MMSEvsK} that when the number of BDs is small, all three array configurations achieve satisfactory estimation performance. However, as the number of BDs increases, the UPA exhibits significantly degraded estimation performance compared to LNA and PNA, owing to its inherently limited angular resolution.

%\begin{figure}[!t]
%  \centering
%  \centerline{\includegraphics[width=2.9in,height=2.1in]{hMSEvsM.eps}}
%  \caption{MSE versus the number of antennas $M$.}
%  \label{MMSEvsM}
%%  \vspace{-0.2cm}
%  \end{figure}
%\begin{figure}[!t]
%  \centering
%  \centerline{\includegraphics[width=2.9in,height=2.1in]{hMSEvstheta.eps}}
%  \caption{MSE versus $\theta_\mathrm{max}$.}
%  \label{MMSEvstheta}
%%  \vspace{-0.2cm}
%  \end{figure}

%Fig. \ref{RBDvsPt} plots the sum rate of BDs $R_{\mathrm{BD}}$ in \eqref{RBDsum} versus ${{\bar{P}}_{d}}$ for ULA and NA under different numbers of BDs.
%It is firstly observed that $R_{\mathrm{BD}}$ increase with ${{\bar{P}}_{d}}$.
%Besides, when ${{\bar{P}}_{d}}$ is sufficiently large, $R_{\mathrm{BD}}$ of NA increase with the number of BDs,
%while that of ULA decrease with the number of BDs.
%This is because that ULA can not estimate AoAs from BDs accurately with limited number of antennas,
%which undermines the sum rates of BDs.
\begin{figure}[!t]
  \centering
  \centerline{\includegraphics[width=2.3in,height=1.7in]{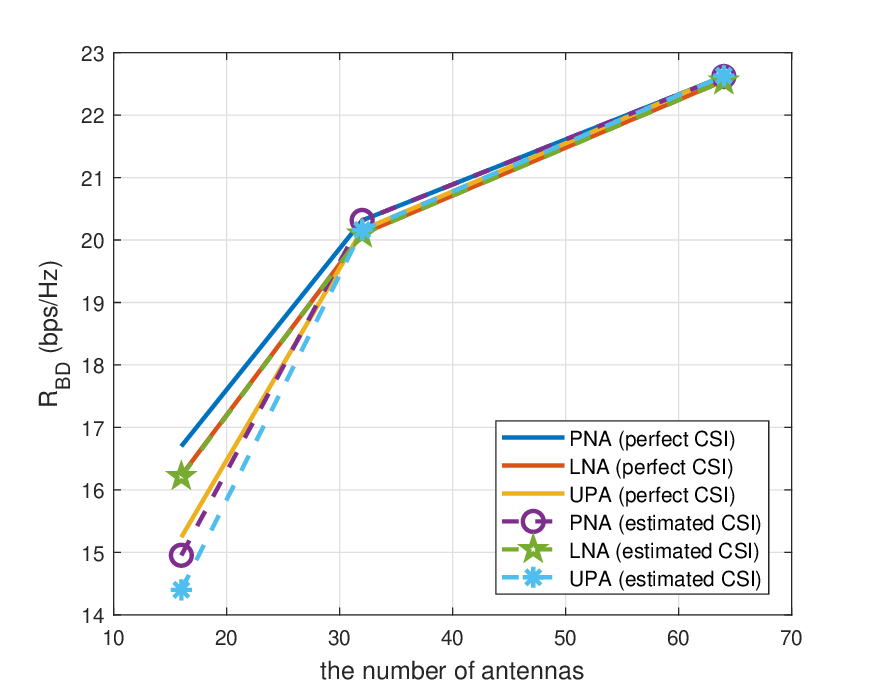}}
  \caption{$R_{\mathrm{BD}}$ versus the number of antennas.}
  \label{RBDvsM}
  \vspace{-0.3cm}
  \end{figure}

\begin{figure}[!t]
  \centering
  \centerline{\includegraphics[width=2.3in,height=1.7in]{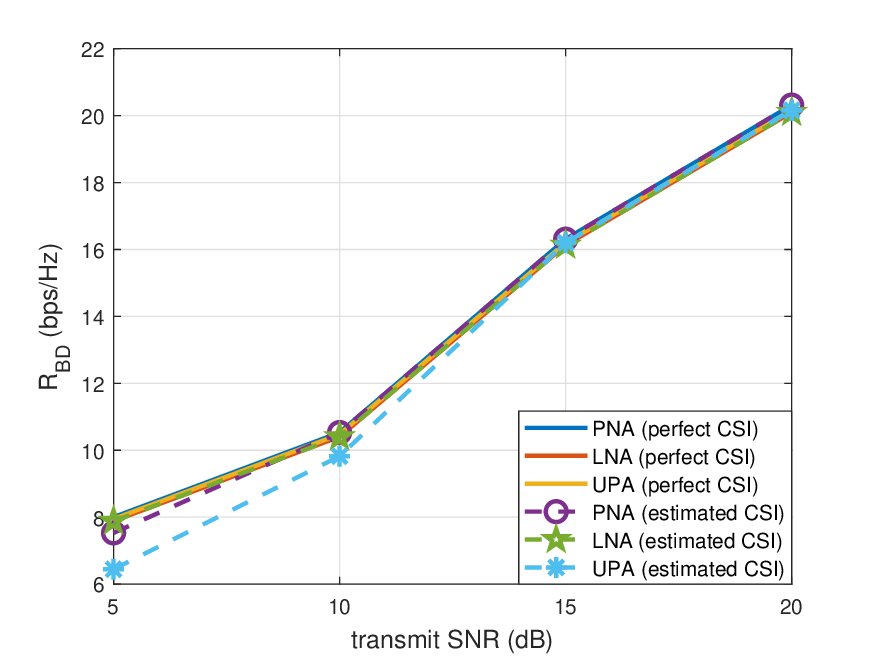}}
  \caption{$R_{\mathrm{BD}}$ versus transmit SNR ${{\bar{P}}_{d}}$.}
  \label{RBDvsPt}
  \vspace{-0.3cm}
  \end{figure}

\begin{figure}[!t]
  \centering
  \centerline{\includegraphics[width=2.3in,height=1.7in]{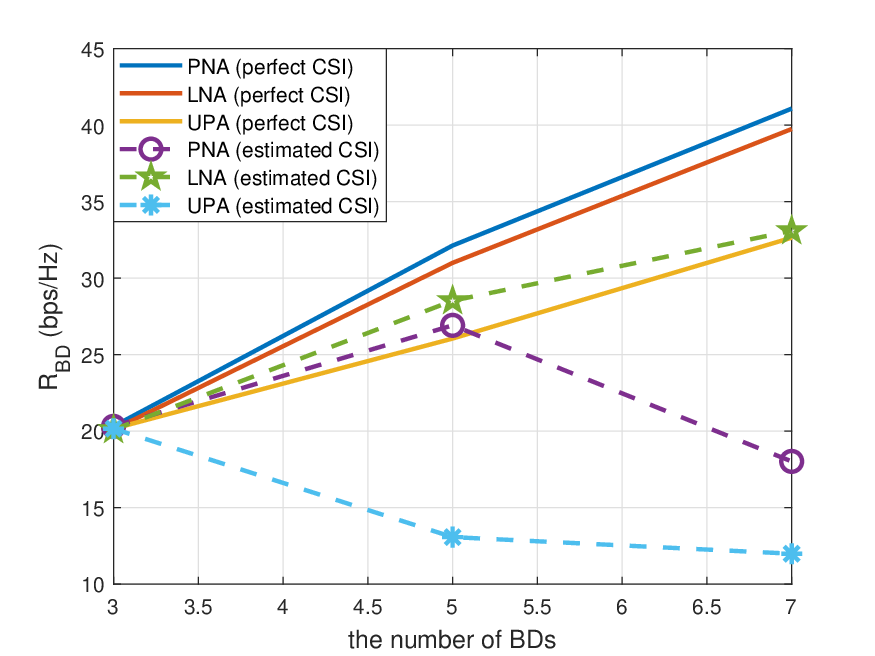}}
  \caption{$R_{\mathrm{BD}}$ versus the number of BDs.}
  \label{RBDvsK}
  \vspace{-0.3cm}
  \end{figure}
Fig. \ref{RBDvsM}, \ref{RBDvsPt} and \ref{RBDvsK} plot the sum rates of BDs based on the estimated CSI and the upper bound values under perfect CSI for PNA, LNA and UPA versus the number of antennas $M$,
transmit SNR ${{\bar{P}}_{t}}$, and the number of BDs $K$, respectively.
It is firstly observed from three figures that under perfect CSI conditions, PNA achieves the highest achievable rate of BDs $R_{\mathrm{BD}}$.
It is also observed from Fig. \ref{RBDvsM} that when the number of antennas is small, e.g., 16 antennas, the actual achievable rates with estimated CSI of PNA and UPA cannot reach the rates of perfect channel estimation, and the actual performance of LNA is better than that of PNA and UPA affected by the estimation accuracy.
When the number of antennas reaches 32, the achievable rates of the three architectures are close to the theoretical upper bound.
It can be observed from Fig. \ref{RBDvsPt} that the gap between the rate with estimated versus perfect CSI narrows as the transmit SNR ${{\bar{P}}_{t}}$ increases.
It can be observed from Fig. \ref{RBDvsK} that when the number of BDs increases, the achievable rates with estimated CSI differ greatly from the rates of perfect channel estimation,
and the achievable rates with estimated CSI of LNA is greater than that of PNA and UPA.
This is consistent with the simulation in Fig. \ref{MMSEvsK}, i.e., when the number of BDs is relatively large, LNA can obtain the best estimation performance.

\section{Conclusion}
In this paper, we consider an IS$^{2}$AC-based SR system with sparse MIMO to achieve both super-resolution 2D sensing and efficient communication for low-altitude UAV swarm.
The achievable rates of UAV swarm and the beam patterns of LNA, PNA and UPA are derived
to evaluate the performance of communication and sensing for IS$^{2}$AC-based SR systems.
To achieve efficient channel estimation for the SR system, we first estimate the 2D DoAs at the BS associated with UAV swarm.
Furthermore, the channel gains of both direct and backscattering links are estimated with a few pilots by matching the paired 2D angles to obtain beamforming gain.
Simulation results are provided to demonstrate the superior performance of LNA and PNA compared to conventional compact arrays.

\begin{appendices}
\section{Proof of Theorem 1}
For PNA, by letting ${{\Delta }_{z}}=0$, the beam pattern on the $y$-axis can be written as
\begin{equation}
  {G_{{{\Delta }_{z}=0}}^{\mathrm{pna}}\left( {{\Delta }_{y}} \right)=\frac{1}{{{M}^{2}}}{{\left| f\left( {{\Delta }_{y}} \right)+g\left( {{\Delta }_{y}} \right)-1 \right|}^{2}}},\label{Gpnay}
\end{equation}
When estimating the smallest local minimum point, the constant term $-1$ in \eqref{Gpnay} can be ignored, which is caused by the shared element at the origin for the sparse and compact UPA, thus resulting
\begin{small}
\begin{equation}
\begin{split}
  & \hat{G}_{{{\Delta }_{z}=0}}^{\mathrm{pna}}\left( {{\Delta }_{y}} \right)=\frac{1}{{{M}^{2}}}{{\left| f\left( {{\Delta }_{y}} \right)+g\left( {{\Delta }_{y}} \right) \right|}^{2}} \\
 & =\frac{1}{{{M}^{2}}}\left( {{f}^{2}}\left( {{\Delta }_{y}} \right)+{{g}^{2}}\left( {{\Delta }_{y}} \right)+2f\left( {{\Delta }_{y}} \right)g\left( {{\Delta }_{y}} \right) \right),
\end{split}
\end{equation}
\end{small}where $f\left( {{\Delta }_{y}} \right)=\frac{M_{2}^{\left( \mathrm{d} \right)}\sin \left( \frac{\pi }{2}\left( 2M_{1}^{\left( \mathrm{d} \right)}+1 \right){{\Delta }_{y}} \right)}{\sin \left( \frac{\pi }{2}{{\Delta }_{y}} \right)}$,
$g\left( {{\Delta }_{y}} \right)=\frac{M_{2}^{\left( \mathrm{s} \right)}\sin (\frac{\pi }{2}(2M_{1}^{\left( \mathrm{d} \right)}+1)(2M_{1}^{\left( \mathrm{s} \right)}+1){{\Delta }_{y}})}{\sin (\frac{\pi }{2}(2M_{1}^{\left( \mathrm{d} \right)}+1){{\Delta }_{y}})}$.
Note that ${{\left| f\left( {{\Delta }_{y}} \right) \right|}^{2}}$ can be
viewed as the beam pattern of a standard compact ULA with $(2M_{1}^{\left( \mathrm{d} \right)}+1)$ antennas,
and ${{\left| g\left( {{\Delta }_{y}} \right) \right|}^{2}}$ can be
viewed as the beam pattern of a sparse ULA with $(2M_{1}^{\left( \mathrm{s} \right)}+1)$ antennas, where the adjacent element spacing is $(2M_{1}^{\left( \mathrm{d} \right)}+1)\frac{\lambda }{2}$.
The smallest local minimum
points of ${{\hat{G}}^{\mathrm{pna}}_{{\Delta }_{z}}}\left( {{\Delta }_{y}} \right)$ is bounded by the first null points of ${{\left| f\left( {{\Delta }_{y}} \right) \right|}^{2}}$ and ${{\left| g\left( {{\Delta }_{y}} \right) \right|}^{2}}$.
For ${{\left| f\left( {{\Delta }_{y}} \right) \right|}^{2}}$, when $\frac{\pi }{2}( 2M_{1}^{\left( \mathrm{d} \right)}+1 ){{\Delta }_{y}}=\pi $,
yielding ${{\Delta }_{y,1}}=\frac{2}{2M_{1}^{\left( \mathrm{d} \right)}+1}$.
For ${{\left| g\left( {{\Delta }_{y}} \right) \right|}^{2}}$, when $\frac{\pi }{2}( 2M_{1}^{\left( \mathrm{d} \right)}+1 )(2M_{1}^{\left( \mathrm{s} \right)}+1){{\Delta }_{y}}=\pi $,
yielding ${{\Delta }_{y,2}}=\frac{2}{( 2M_{1}^{\left( \mathrm{d} \right)}+1 )(2M_{1}^{\left( \mathrm{s} \right)}+1)}$.

First, since $f\left( {{\Delta }_{y}} \right)$ and $g\left( {{\Delta }_{y}} \right)$ decreases monotonically for $0<{{\Delta }_{y}}<{{\Delta }_{y,2}}$,
we have ${{\Delta }_{y,\min }}>{{\Delta }_{y,2}}$.

Next, for ${{\Delta }_{y}}={{\Delta }_{y,1}}$, the derivative of $\hat{G}_{{{\Delta }_{z}}}^{\mathrm{pna}}\left( {{\Delta }_{y}} \right)$
can be calculated as
\begin{small}
\begin{equation}
{\frac{\mathrm{d}\big( {{\hat{G}}^{\mathrm{pna}}_{{\Delta }_{z}=0}}\left( {{\Delta }_{y}} \right) \big)}{\mathrm{d}\left( {{\Delta }_{y}} \right)}=\frac{2g\left( {{\Delta }_{y}} \right)}{{{M}^{2}}}\Big( \frac{\mathrm{d}\left( g\left( {{\Delta }_{y}} \right) \right)}{\mathrm{d}\left( {{\Delta }_{y}} \right)}+\frac{\mathrm{d}\left( f\left( {{\Delta }_{y}} \right) \right)}{\mathrm{d}\left( {{\Delta }_{y}} \right)} \Big)}.
\end{equation}
\end{small}
Due to the fact that $g\left( {{\Delta }_{y}} \right)$ reaches its local value when ${{\Delta }_{y}}={{\Delta }_{y,1}}$,
we have $g\left( {{\Delta }_{y}} \right)\frac{\text{d}\left( g\left( {{\Delta }_{y}} \right) \right)}{\text{d}\left( {{\Delta }_{y}} \right)}>0$ for ${{\Delta }_{y,\min }}\to {{{\Delta }_{y,1}}^{-}}$,
and $g\left( {{\Delta }_{y}} \right)\frac{\text{d}\left( g\left( {{\Delta }_{y}} \right) \right)}{\text{d}\left( {{\Delta }_{y}} \right)}<0$ for ${{\Delta }_{y,\min }}\to {{{\Delta }_{y,1}}^{+}}$.
Furthermore, since the slope of $g\left( {{\Delta }_{y}} \right)$ is steeper than that of $f\left( {{\Delta }_{y}} \right)$,
we have $\frac{\mathrm{d}\left( {{\hat{G}}^{\mathrm{pna}}_{{\Delta }_{z}=0}}\left( {{\Delta }_{y}} \right) \right)}{\mathrm{d}\left( {{\Delta }_{y}} \right)}>0$ for ${{\Delta }_{y,\min }}\to {{{\Delta }_{y,1}}^{-}}$,
and $\frac{\mathrm{d}\left( {{\hat{G}}^{\mathrm{pna}}_{{\Delta }_{z}=0}}\left( {{\Delta }_{y}} \right) \right)}{\mathrm{d}\left( {{\Delta }_{y}} \right)}<0$ for ${{\Delta }_{y,\min }}\to {{{\Delta }_{y,1}}^{+}}$.
Therefore, ${\hat{G}}^{\mathrm{pna}}_{{{\Delta }_{z}=0}}\left( {{\Delta }_{y}} \right)$ obtain its local maximum value for ${{\Delta }_{y}}={{\Delta }_{y,1}}$.
The smallest local minimum points for the positive semi-axis of $y$ is upper bounded by ${{\Delta }_{y,\min }}<{{\Delta }_{y,1}}$.

\begin{figure}[!t]
  \centering
  \centerline{\includegraphics[width=2.3in,height=1.7in]{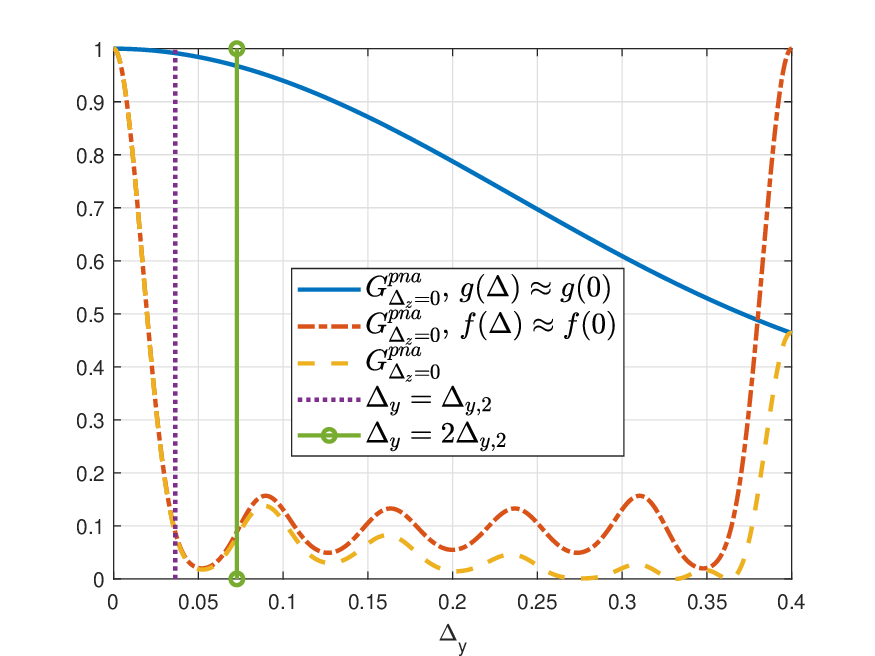}}
  \caption{Trajectory of $G_{{{\Delta }_{z}=0}}^{\mathrm{pna}}\left( {{\Delta }_{y}} \right)$ with $M_{1}^{\left( \mathrm{d}\right)}=2,M_{2}^{\left( \mathrm{d}\right)}=3,M_{1}^{\left( \mathrm{s}\right)}=5,M_{2}^{\left( \mathrm{s}\right)}=3$.}
  \label{GPNAy}
%  \vspace{-0.3cm}
  \end{figure}

Then the main lobe width on the $y$-axis is bounded by
\begin{small}
\begin{equation}
{\frac{4}{(2M_{1}^{\left( \mathrm{d} \right)}+1)(2M_{1}^{\left( \mathrm{s} \right)}+1)}<BW_y<\frac{4}{2M_{1}^{\left( \mathrm{d} \right)}+1}}.
\end{equation}
\end{small}
Specifically, by numerical simulation, we found that when $M_{1}^{\left( \text{s} \right)}\ge \frac{M_{2}^{\left( \text{d} \right)}}{M_{2}^{\left( \text{s} \right)}}\big( 2M_{1}^{\left( \text{d} \right)}+1 \big)$, $g\left( {{\Delta }_{y}} \right)$ varies much faster than $f\left( {{\Delta }_{y}} \right)$, which causes the smallest local
minimum points to be very close to ${{\Delta }_{y,2}}$,
and be smaller than $2{{\Delta }_{y,2}}$.
That is to say, the main lobe width on the $y$-axis will approach $\frac{4}{(2M_{1}^{\left( \mathrm{d} \right)}+1)(2M_{1}^{\left( \mathrm{s} \right)}+1)}$ and further upper bounded by
$BW_y<\frac{8}{(2M_{1}^{\left( \mathrm{d} \right)}+1)(2M_{1}^{\left( \mathrm{s} \right)}+1)}$.
An example of the trajectory of $G_{{{\Delta }_{z}=0}}^{\mathrm{pna}}\left( {{\Delta }_{y}} \right)$ is shown in Fig. \ref{GPNAy} with $M_{1}^{\left( \mathrm{d}\right)}=2,M_{2}^{\left( \mathrm{d}\right)}=3,M_{1}^{\left( \mathrm{s}\right)}=5,M_{2}^{\left( \mathrm{s}\right)}=3$.

The proof of Theorem 1 is completed.
\section{Proof of Theorem 2}
By letting ${{\Delta }_{y}}=0$, the beam pattern on the $z$-axis can be written as
\begin{small}
\begin{equation}
 { G_{{{\Delta }_{y}}=0}^{\text{pna}}\left( {{\Delta }_{z}} \right)=a{{\bigg| p\left( {{\Delta }_{z}} \right)+b{{e}^{jr\left( {{\Delta }_{z}} \right)}}q\left( {{\Delta }_{z}} \right)-\frac{1}{2M_{1}^{\left( \text{d} \right)}+1} \bigg|}^{2}}},\label{Gpnaz}
\end{equation}
\end{small}where $a=\frac{{{\left( 2M_{1}^{\left( \text{d} \right)}+1 \right)}^{2}}}{{{M}^{2}}}$,
$b=\frac{2M_{1}^{\left( \text{s} \right)}+1}{2M_{1}^{\left( \text{d} \right)}+1}$,
$p\left( {{\Delta }_{z}} \right)=\frac{\sin \left( \frac{\pi }{2}M_{2}^{\left( \text{d} \right)}{{\Delta }_{z}} \right)}{\sin \left( \frac{\pi }{2}{{\Delta }_{z}} \right)}$,
$q\left( {{\Delta }_{z}} \right)=\frac{\sin (\frac{\pi }{2}M_{2}^{\left( \text{d} \right)}M_{2}^{\left( \text{s} \right)}{{\Delta }_{z}})}{\sin (\frac{\pi }{2}M_{2}^{\left( \text{d} \right)}{{\Delta }_{z}})}$,
$r\left( {{\Delta }_{z}} \right)=\frac{\pi }{2}\big( M_{1}^{\left( \text{d} \right)}+1 \big)\big( M_{2}^{\left( \text{s} \right)}-1 \big){{\Delta }_{z}}$.
When estimating the smallest local minimum point, the constant term $-\frac{1}{2M_{1}^{\left( \text{d} \right)}+1}$ in \eqref{Gpnaz} can be ignored, which is caused by the shared element at the origin for the sparse and compact UPA, thus resulting
\begin{small}
\begin{equation}
\begin{split}
  & \hat{G}_{{{\Delta }_{y}}=0}^{\text{pna}}\left( {{\Delta }_{z}} \right)=a{{\left| p\left( {{\Delta }_{z}} \right)+b{{e}^{jr\left( {{\Delta }_{z}} \right)}}q\left( {{\Delta }_{z}} \right) \right|}^{2}} \\
 & =a\left( {{p}^{2}}\left( {{\Delta }_{z}} \right)+{{b}^{2}}{{q}^{2}}\left( {{\Delta }_{z}} \right)+2b\cos \left( r\left( {{\Delta }_{z}} \right) \right)p\left( {{\Delta }_{z}} \right)q\left( {{\Delta }_{z}} \right) \right).
\end{split}
\end{equation}
\end{small}

Since the smallest local minimum point for the positive semi-axis of $z$ can not be solved in
closed-form, we first calculate the first null points of $p\left( {{\Delta }_{z}} \right)$, $q\left( {{\Delta }_{z}} \right)$ and $\cos\left(r\left( {{\Delta }_{z}} \right)\right)$.
For $p\left( {{\Delta }_{z}} \right)$, when $\frac{\pi }{2} M_{2}^{\left( \mathrm{d} \right)}{{\Delta }_{z}}=\pi $,
yielding ${{\Delta }_{z,1}}=\frac{2}{M_{2}^{\left( \mathrm{d} \right)}}$.
For $q\left( {{\Delta }_{z}} \right)$, when $\frac{\pi }{2}M_{2}^{\left( \mathrm{d}\right)}M_{2}^{\left( \mathrm{s}\right)}{{\Delta }_{z}}=\pi $,
yielding ${{\Delta }_{z,2}}=\frac{2}{M_{2}^{\left( \mathrm{d} \right)}M_{2}^{\left( \mathrm{s}\right)}}$.
For $\cos\left(r\left( {{\Delta }_{z}} \right)\right)$,
when $\frac{\pi }{2}\big( M_{1}^{\left( \text{d} \right)}+1 \big)\big( M_{2}^{\left( \text{s} \right)}-1 \big){{\Delta }_{z}}=\frac{\pi }{2}$,
yielding ${{\Delta }_{z,3}}=\frac{1}{\left( M_{1}^{\left( \text{d} \right)}+1 \right)\left( M_{2}^{\left( \text{s} \right)}-1 \right)}$.

Since $p\left( {{\Delta }_{z}} \right)$, $q\left( {{\Delta }_{z}} \right)$ and $\cos\left(r\left( {{\Delta }_{z}} \right)\right)$ decrease monotonically for $0<{{\Delta }_{z}}<\min \left\{ {{\Delta }_{z,2}},{{\Delta }_{z,3}} \right\}$,
we have ${{\Delta }_{z,\min }}>\min \left\{ {{\Delta }_{z,2}},{{\Delta }_{z,3}} \right\}$.

Furthermore, for ${{\Delta }_{z}}={{\Delta }_{z,1}}$, the derivative of $\hat{G}_{{{\Delta }_{y}}=0}^{\text{pna}}\left( {{\Delta }_{z}} \right)$
can be calculated as
\begin{small}
\begin{equation}
\begin{split}
& \frac{\text{d}\Big( \hat{G}_{{{\Delta }_{y}}=0}^{\text{pna}}\left( {{\Delta }_{z}} \right) \Big)}{\text{d}\left( {{\Delta }_{y}} \right)}\\
& =2a{{b}^{2}}q\left( {{\Delta }_{z}} \right)\Big( \frac{\text{d}\left( q\left( {{\Delta }_{z}} \right) \right)}{\text{d}\left( {{\Delta }_{y}} \right)}+\frac{\cos \left( r\left( {{\Delta }_{z}} \right) \right)}{b}\frac{\text{d}\left( p\left( {{\Delta }_{z}} \right) \right)}{\text{d}\left( {{\Delta }_{y}} \right)} \Big).
\end{split}
\end{equation}
\end{small}
When $b\ge 1$, i.e., $M_{1}^{\left( \text{s} \right)}\ge M_{1}^{\left( \text{d} \right)}$,
$\frac{\cos \left( r\left( {{\Delta }_{z}} \right) \right)}{b}\le 1$.
Since $q\left( {{\Delta }_{y}} \right)$ changes faster than $p\left( {{\Delta }_{y}} \right)$,
we have $\left| \frac{\text{d}\left( q\left( {{\Delta }_{z}} \right) \right)}{\text{d}\left( {{\Delta }_{y}} \right)} \right|\ge \left| \frac{\text{d}\left( p\left( {{\Delta }_{z}} \right) \right)}{\text{d}\left( {{\Delta }_{y}} \right)} \right|$.
Following a similar derivation as above where ${\hat{G}}^{\mathrm{pna}}_{{{\Delta }_{z}=0}}\left( {{\Delta }_{y}} \right)$ obtains its first local maximum value for ${{\Delta }_{y}}={{\Delta }_{y,1}}$,
$\hat{G}_{{{\Delta }_{y}}=0}^{\text{pna}}\left( {{\Delta }_{z}} \right)$ can achieve the first local maximum value when ${{\Delta }_{z}}={{\Delta }_{z,1}}$.
Therefore, the smallest local minimum points for the positive semi-axis of $z$ is upper bounded by ${{\Delta }_{z,\min }}<{{\Delta }_{z,1}}$,
and the main lobe width on the $z$-axis is bounded by
\begin{small}
\begin{equation}
{2\min \left\{ {{\Delta }_{z,2}},{{\Delta }_{z,3}} \right\}<BW_z<\frac{4}{M_{2}^{\left( \mathrm{d} \right)}}}.
\end{equation}
\end{small}

\begin{figure}[!t]
  \centering
  \centerline{\includegraphics[width=2.3in,height=1.7in]{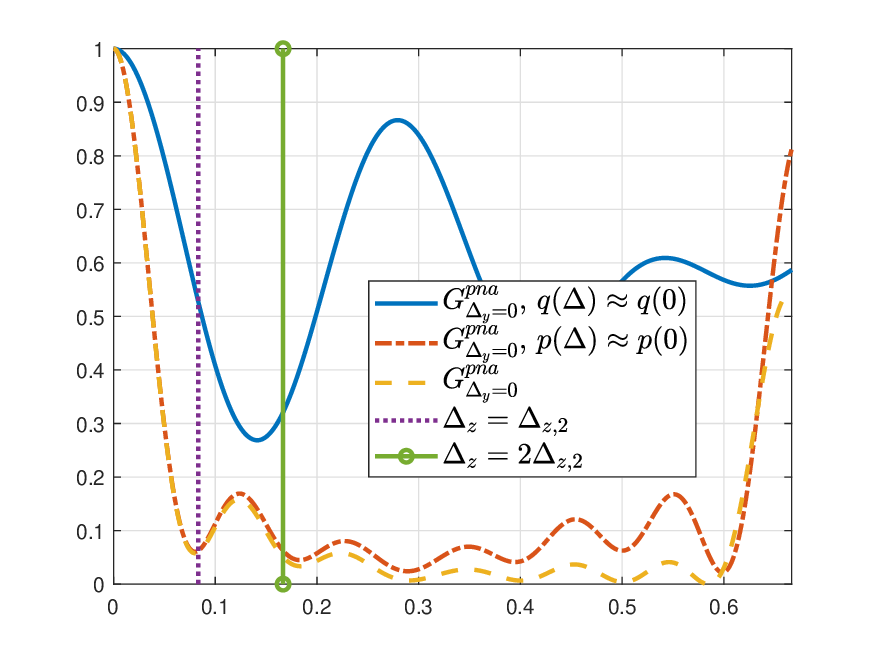}}
  \caption{Trajectory of $G_{{{\Delta }_{y}=0}}^{\mathrm{pna}}\left( {{\Delta }_{z}} \right)$ with $M_{1}^{\left( \mathrm{d}\right)}=1,M_{2}^{\left( \mathrm{d}\right)}=3,M_{1}^{\left( \mathrm{s}\right)}=1,M_{2}^{\left( \mathrm{s}\right)}=8$.}
  \label{GPNAz}
%  \vspace{-0.3cm}
  \end{figure}

Specifically, by numerical simulation, we found that when $M_{2}^{\left( \text{s} \right)}\ge \frac{2M_{1}^{\left( \text{d} \right)}+1}{2M_{1}^{\left( \text{s} \right)}+1}\big( 3M_{2}^{\left( \text{d} \right)}-1 \big)$, $q\left( {{\Delta }_{z}} \right)$ varies much faster than $p\left( {{\Delta }_{z}} \right)$, which causes the smallest local
minimum points to be very close to ${{\Delta }_{z,2}}$,
and the derivative of $\hat{G}_{{{\Delta }_{y}}=0}^{\text{pna}}\left( {{\Delta }_{z}} \right)$ is greater than 0 at $2{{\Delta }_{z,2}}$.
That is to say, the main lobe width on the $z$-axis is very close to $\frac{4}{M_{2}^{\left( \mathrm{d} \right)}M_{2}^{\left( \mathrm{s}\right)}}$ and further upper bounded by
$BW_z<\frac{8}{M_{2}^{\left( \mathrm{d} \right)}M_{2}^{\left( \mathrm{s}\right)}}$.
An example of the trajectory of $G_{{{\Delta }_{y}=0}}^{\mathrm{pna}}\left( {{\Delta }_{z}} \right)$ is shown in Fig. \ref{GPNAz} with $M_{1}^{\left( \mathrm{d}\right)}=1,M_{2}^{\left( \mathrm{d}\right)}=3,M_{1}^{\left( \mathrm{s}\right)}=1,M_{2}^{\left( \mathrm{s}\right)}=8$.

The proof of Theorem 2 is thus completed.
\section{Proof of Theorem 3}
For LNA, by letting ${{\Delta }_{z}}=0$, the beam pattern on the $y$-axis can be written as
\begin{small}
\begin{equation}
  {G_{{{\Delta }_{z}}=0}^{\text{lna}}\left( {{\Delta }_{y}} \right)\!=\!\frac{1}{{{M}^{2}}}{{\left| h\left( {{\Delta }_{y}} \right)\!+\!{{e}^{jw\left( {{\Delta }_{y}} \right)}}s\left( {{\Delta }_{y}} \right)\!+\!{{e}^{jv\left( {{\Delta }_{y}} \right)}}{{M}_{z}}-1 \right|}^{2}}}.\label{GLNAy}
\end{equation}
\end{small}
When estimating the smallest local minimum point, the constant term $-1$ in \eqref{GLNAy} can be ignored,
thus resulting
\begin{small}
\begin{equation}
\begin{split}
  & G_{{{\Delta }_{z}}=0}^{\text{lna}}\left( {{\Delta }_{y}} \right)=\frac{1}{{{M}^{2}}}{{\left| h\left( {{\Delta }_{y}} \right)+{{e}^{jw\left( {{\Delta }_{y}} \right)}}s\left( {{\Delta }_{y}} \right)+{{e}^{jv\left( {{\Delta }_{y}} \right)}}{{M}_{z}} \right|}^{2}} \\
 & =\frac{1}{{{M}^{2}}}\left( {{h}^{2}}\left( {{\Delta }_{y}} \right)+{{s}^{2}}\left( {{\Delta }_{y}} \right)+M_{z}^{2}+2\cos \left( w\left( {{\Delta }_{y}} \right) \right)h\left( {{\Delta }_{y}} \right)s\left( {{\Delta }_{y}} \right) \right) \\
 & \!+\!\frac{1}{{{M}^{2}}}\left( 2{{M}_{z}}\cos \left( v\left( {{\Delta }_{y}} \right) \right)h\left( {{\Delta }_{y}} \right)\!+\!2{{M}_{z}}\cos \left( w\left( {{\Delta }_{y}} \right)\!-\!v\left( {{\Delta }_{y}} \right) \right)s\left( {{\Delta }_{y}} \right) \right),
\end{split}
\end{equation}
\end{small}where $h\left( {{\Delta }_{y}} \right)=\frac{\sin \left( \frac{\pi }{2}{{M}_{y,1}}{{\Delta }_{y}} \right)}{\sin \left( \frac{\pi }{2}{{\Delta }_{y}} \right)}$,
$s\left( {{\Delta }_{y}} \right)=\frac{\sin \left( \frac{\pi }{2}\left( {{M}_{y,1}}+1 \right){{M}_{y,2}}{{\Delta }_{y}} \right)}{\sin \left( \frac{\pi }{2}\left( {{M}_{y,1}}+1 \right){{\Delta }_{y}} \right)}$,
$w\left( {{\Delta }_{y}} \right)=\frac{\pi }{2}\left( {{M}_{y,1}}+1 \right){{M}_{y,2}}{{\Delta }_{y}}$,
$v\left( {{\Delta }_{y}} \right)=-\frac{\pi }{2}\left( {{M}_{y,1}}-1 \right){{\Delta }_{y}}$.
We first calculate the first null points of $h\left( {{\Delta }_{y}} \right)$, $s\left( {{\Delta }_{y}} \right)$, $\cos\left(w\left( {{\Delta }_{y}} \right)\right)$, and $\cos\left(v\left( {{\Delta }_{y}} \right)\right)$.
For $h\left( {{\Delta }_{y}} \right)$, when $\frac{\pi }{2} M_{y,1}{{\Delta }_{y}}=\pi $,
yielding ${{\Delta }_{y,3}}=\frac{2}{M_{y,1}}$.
For $s\left( {{\Delta }_{y}} \right)$, when $\frac{\pi }{2}\left(M_{y,1}+1\right)M_{y,2}{{\Delta }_{y}}=\pi $,
yielding ${{\Delta }_{y,4}}=\frac{2}{\left(M_{y,1}+1\right)M_{y,2}}$.
For $\cos\left(w\left( {{\Delta }_{y}} \right)\right)$,
when $\frac{\pi }{2}\left( M_{y,1}+1 \right)M_{y,2}{{\Delta }_{y}}=\frac{\pi }{2}$,
yielding ${{\Delta }_{y,5}}=\frac{1}{\left( M_{y,1}+1 \right)M_{y,2}}$.
For $\cos\left(v\left( {{\Delta }_{y}} \right)\right)$,
when $\frac{\pi }{2}\left( M_{y,1}-1 \right){{\Delta }_{y}}=\frac{\pi }{2}$,
yielding ${{\Delta }_{y,6}}=\frac{1}{\left( M_{y,1}-1 \right)}$.
Without loss of generality, assume ${{M}_{y,1}}={{M}_{y,2}}\ge 2$,
then we have ${{\Delta }_{y,5}}<{{\Delta }_{y,4}}<{{\Delta }_{y,6}}\le{{\Delta }_{y,3}}$.
Since $h\left( {{\Delta }_{y}} \right)$, $s\left( {{\Delta }_{y}} \right)$, $\cos\left(w\left( {{\Delta }_{y}} \right)\right)$ and $\cos\left(v\left( {{\Delta }_{y}} \right)\right)$ decrease monotonically for $0<{{\Delta }_{y}}<{{\Delta }_{y,5}}$,
we have ${{\Delta }_{y,\min }}>{{\Delta }_{y,5}}$.

Furthermore, for ${{\Delta }_{y}}={{\Delta }_{y,3}}$, the derivative of $G_{{{\Delta }_{z}}=0}^{\text{lna}}\left( {{\Delta }_{y}} \right)$
can be calculated as
\begin{small}
\begin{equation}
\begin{split}
  & \frac{\text{d}\left( G_{{{\Delta }_{z}}=0}^{\text{lna}}\left( {{\Delta }_{y}} \right) \right)}{\text{d}\left( {{\Delta }_{y}} \right)}= \\
 & \frac{2{{M}_{z}}}{{{M}^{2}}}\Big( \cos \left( v\left( {{\Delta }_{y}} \right) \right)\frac{\text{d}\left( h\left( {{\Delta }_{y}} \right) \right)}{\text{d}\left( {{\Delta }_{y}} \right)}\!+\!\cos \left( w\left( {{\Delta }_{y}} \right)\!-\!v\left( {{\Delta }_{y}} \right) \right)\frac{\text{d}\left( s\left( {{\Delta }_{y}} \right) \right)}{\text{d}\left( {{\Delta }_{y}} \right)} \Big),
\end{split}
\end{equation}
\end{small}where $\cos \left( v\left( {{\Delta }_{y}} \right) \right)\le 0$, $\frac{\text{d}\left( h\left( {{\Delta }_{y}} \right) \right)}{\text{d}\left( {{\Delta }_{y}} \right)}\le 0$,
$\cos \left( w\left( {{\Delta }_{y}} \right)-v\left( {{\Delta }_{y}} \right) \right)\frac{\text{d}\left( s\left( {{\Delta }_{y}} \right) \right)}{\text{d}\left( {{\Delta }_{y}} \right)}\ge 0$.
Therefore, $\frac{\text{d}\left( G_{{{\Delta }_{z}}=0}^{\text{lna}}\left( {{\Delta }_{y}} \right) \right)}{\text{d}\left( {{\Delta }_{y}} \right)}\le 0$ for ${{\Delta }_{y}}={{\Delta }_{y,3}}$,
which results in ${{\Delta }_{y,\min }}>{{\Delta }_{y,3}}$.
Then the main lobe width on the $y$-axis is bounded by
\begin{equation}
{\frac{2}{{\left( M_{y,1}+1 \right)M_{y,2}}}<BW_y<\frac{4}{M_{y,1}}}.
\end{equation}
Specifically, by numerical simulation, we found that when ${{M}_{y,2}}\ge 3\left( {{M}_{y,1}}+1 \right)$, the variation of $s\left( {{\Delta }_{y}} \right)$ is the dominant factor of $G_{{{\Delta }_{z}}=0}^{\text{lna}}\left( {{\Delta }_{y}} \right)$, which causes the smallest local
minimum points to be very close to and upper bounded by ${{\Delta }_{y,4}}$.
That is to say, the main lobe width on the $y$-axis is further upper bounded by
$BW_y<\frac{4}{\left(M_{y,1}+1\right)M_{y,2}}$.
An example of the trajectory of $G_{{{\Delta }_{z}=0}}^{\mathrm{lna}}\left( {{\Delta }_{y}} \right)$ is shown in Fig. \ref{GLNAy} with $M_{y,1}=1,M_{y,2}=6,M_{z,1}=1,M_{z,2}=6$.
\begin{figure}[!t]
  \centering
  \centerline{\includegraphics[width=2.3in,height=1.7in]{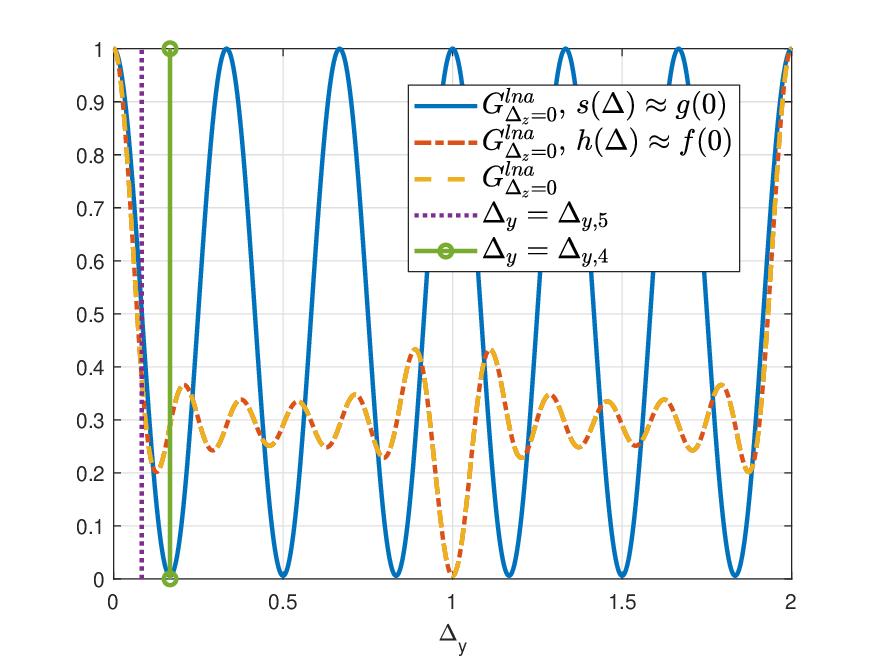}}
  \caption{Trajectory of $G_{{{\Delta }_{z}=0}}^{\mathrm{lna}}\left( {{\Delta }_{y}} \right)$ with $M_{y,1}=1,M_{y,2}=6,M_{z,1}=1,M_{z,2}=6$.}
  \label{GLNAy}
%  \vspace{-0.3cm}
  \end{figure}

The proof of Theorem 3 is completed.
\end{appendices}


\begin{thebibliography}{99}
\bibitem{UAV1}
Y. Zeng, Q. Wu, and R. Zhang, ``Accessing from the sky: A tutorial on UAV communications for 5G and beyond," \emph{Proc. IEEE}, vol. 107, no. 12, pp. 2327--2375, Dec. 2019.
\bibitem{UAV2}
Y. Zeng and R. Zhang, ``Energy-efficient UAV communication with trajectory optimization," \emph{IEEE Trans. Wireless Commun.}, vol. 16, no. 6, pp. 3747--3760, Jun. 2017.
\bibitem{UAV3}
L. Gupta, R. Jain, and G. Vaszkun, ``Survey of important issues in UAV communication networks," \emph{IEEE Commun. Surv. Tutorials}, vol. 18, no. 2, pp. 1123--1152, 2016.
\bibitem{vtc}
J. Xu, Y. Zeng, F. Yang and Y. Chen, ``Integrated super-resolution sensing
and communication with sparse MIMO for symbiotic radio," \emph{in Proc. 2025 IEEE Veh. Technol. Conf. (VTC)}, 2025, pp. 1--6.
\bibitem{UAVswarm1}
L. Ma, B. Lin, W. Zhang, J. Tao, X. Zhu, and H. Chen, ``A survey of research on the distributed cooperation method of the UAV swarm based on swarm intelligence," \emph{in Proc. IEEE Int. Conf. Softw. Eng. Service Sci. (ICSESS)}, Beijing, China, 2022, pp. 305--309.
\bibitem{UAVswarm2}
M. Campion, P. Ranganathan, and S. Faruque, ``UAV swarm communication and control architectures: a review," \emph{J. Unmanned Veh. Syst.}, vol. 7, no. 2, pp. 93--106, 2018.
\bibitem{UAVswarm3}
S. Javed et al., ``State-of-the-art and future research challenges in UAV swarms," \emph{IEEE Internet Things J.}, vol. 11, no. 11, pp. 19023--19045, Jun. 2024.
%\bibitem{IMT2030}
%ITU-R, DRAFT NEW RECOMMENDATION, ``Framework and overall objectives of the future development of IMT for 2030 and beyond," June 2023.
%\bibitem{liufan}
%F. Liu, Y. Cui, C. Masouros, J. Xu, T. X. Han, Y. C. Eldar, and S. Buzzi, ``Integrated sensing and communications: Toward dual-functional wireless networks for 6G and beyond,"
%\emph{IEEE J. Select. Areas Commun.}, vol. 40, no. 6, pp. 1728--1767, Jun. 2022.
\bibitem{SR1}
Y.-C. Liang, Q.~Zhang, E.~G.~Larsson and G.~Y.~Li, ``Symbiotic radio: Cognitive backscattering communications for future wireless networks," \emph{IEEE Trans. Cognit. Commun. Netw.}, vol.~6, no.~4, pp. 1242--1255, Dec.~2020.
\bibitem{SR2}
R. Long, Y. -C. Liang, H. Guo, G. Yang and R. Zhang, ``Symbiotic radio: A new communication paradigm for passive Internet of Things," \emph{IEEE Internet Things J.}, vol. 7, no. 2, pp. 1350--1363, Feb. 2020.
\bibitem{SR3}
Y. -C. Liang, R. Long, Q. Zhang and D. Niyato, "Symbiotic communications: Where marconi meets darwin," \emph{IEEE Wireless Commun.}, vol. 29, no. 1, pp. 144--150, Feb. 2022.
\bibitem{SR4}
J. Xu, Z. Dai and Y. Zeng, ``MIMO symbiotic radio with massive backscatter devices: Asymptotic analysis and precoding optimization," \emph{IEEE Trans. Commun.}, vol. 71, no. 9, pp. 5487--5502, Sept. 2023.
\bibitem{SR5}
Z. Dai, R. Li, J. Xu, Y. Zeng and S. Jin, ``Rate-region characterization and channel estimation for cell-free symbiotic radio communications," \emph{IEEE Trans. Commun.}, vol. 71, no. 2, pp. 674--687, Feb. 2023.
\bibitem{zhangchaoyue}
C. Zhang, Z. Zhou, H. Wang and Y. Zeng, ``Integrated Super-Resolution
Sensing and Communication with 5G NR Waveform: Signal Processing
with Uneven CPs and Experiments: (Invited Paper)," \emph{in Proc. Int. Symp.
Model. and Opt. Mobile, Ad Hoc, and Wireless Networks (WiOpt)}, 2023,
pp. 681--688.
\bibitem{ISSACJournal}
J. Xu, H. Wang, Y. Zeng, X. Xu, Q. Wu, F. Yang, Y. Chen, and A. Jamalipour, ``Efficient channel estimation for millimeter wave and
terahertz systems enabled by integrated super-resolution sensing and communication," \emph{arXiv preprint arXiv:2407.20607}, 2024.
%\bibitem{CR1}
%J.~Mitola and G.~Q.~Maguire, ``Cognitive radio: Making software radios more personal," \emph{IEEE Pers. Commun.}, vol.~6, no.~4, pp.~13--18, Aug.~1999.
%\bibitem{CR2}
%S. Haykin, ``Cognitive radio: brain-empowered wireless communications," \emph{IEEE J. Sel. Areas Commun.}, vol. 23, no. 2, pp. 201--220, 2005.
%\bibitem{CR3}
%Y.-C. Liang, K.-C. Chen, G. Y. Li, and P. Mahonen, ``Cognitive radio networking and communications: An overview," \emph{IEEE Trans.
%Veh. Tech.}, vol. 60, no. 7, pp. 3386--3407, 2011.
%\bibitem{SRISAC1}
%D. Galappaththige, C. Tellambura and A. Maaref, ``Integrated sensing and backscatter communication," \emph{IEEE Wireless Commun. Lett.}, vol.
%12, no. 12, pp. 2043--2047, Dec. 2023.
%\bibitem{SRISAC2}
%Q. Tao, X. Hu, S. Zhang, and C. Zhong, ``Integrated sensing and communication for
%symbiotic radio systems in mobile scenarios," \emph{IEEE Trans. Wirelese Commun.}, vol. 23, no. 9, pp. 11213--11225, Sept. 2024.
%\bibitem{SRISAC3}
%X. Wang, Z. Fei and Q. Wu, ``Integrated sensing and communication for RIS-assisted backscatter systems," \emph{
%IEEE Internet of Things J.}, vol. 10, no. 15, pp. 13716--13726, Aug. 2023.
%\bibitem{SRISAC4}
%Q. Tao, C. Huang and X. Chen, ``Integrated sensing and communication for
%symbiotic radio with multiple IoT devices," \emph{IEEE Commun. Lett.}, vol. 28, no. 8, pp. 1820--1824, Aug. 2024.
\bibitem{lixinrui}
X. Li, H. Min, Y. Zeng, S. Jin, L. Dai, Y. Yuan, and R. Zhang, ``Sparse
MIMO for ISAC: New opportunities and challenges," accepted by \emph{IEEE Wireless Commun.}, \emph{arXiv preprint arXiv:2406.12270}, 2024.
\bibitem{wanghuizhi}
H. Wang, C. Feng, Y. Zeng, S. Jin, C. Yuen, B. Clerckx, and R. Zhang, ``Enhancing spatial multiplexing and interference suppression for near- and far-Field
communications with sparse MIMO," \emph{arXiv preprint arXiv:2408.01956}, 2024.
\bibitem{MRA}
C.-Y. Chen and P. P. Vaidyanathan, ``Minimum redundancy MIMO
radars," \emph{in Proc. IEEE Int. Symp. Circuits. Syst. (ISCAS)}, 2008, pp. 45--48.
\bibitem{NA1}
P. Pal and P. P. Vaidyanathan, ``Nested arrays: A novel approach to
array processing with enhanced degrees of freedom," \emph{IEEE Trans. Signal
Processing}, vol. 58, no. 8, pp. 4167--4181, Aug. 2010.
\bibitem{MoA}
X. Li, Z. Dong, Y. Zeng, S. Jin, and R. Zhang, ``Multi-user modular
XL-MIMO communications: Near-field beam focusing pattern and
user grouping," \emph{IEEE Trans. Wireless Commun.}, vol. 23, no. 10, pp. 13766--13781, Oct. 2024.
\bibitem{CPA}
P. P. Vaidyanathan and P. Pal, ``Sparse sensing with co-prime samplers
and arrays," \emph{IEEE Trans. Signal Process.}, vol. 59, no. 2, pp. 573--586,
2011.
\bibitem{2Dsparseaarray}
I. Aboumahmoud, A. Muqaibel, M. Alhassoun and S. Alawsh, ``A review of sparse sensor arrays for two-dimensional direction-of-arrival estimation," \emph{IEEE Access}, vol. 9, pp. 92999--93017, 2021.
\bibitem{parallel1}
J. Li, X. Zhang, and H. Chen, ``Improved two-dimensional DOA estimation algorithm for two-parallel
uniform linear arrays using propagator method," \emph{Signal Process.}, vol. 92, no. 12, pp. 3032--3038, Dec. 2012.
\bibitem{parallel2}
F. Sun, P. Lan, B. Gao, and G. Zhang, ``An efficient dictionary learning-based 2-D DOA estimation without pair matching for co-prime parallel
arrays," \emph{IEEE Access}, vol. 6, pp. 8510--8518, 2018.
\bibitem{parallel3}
Z. Zheng, Y. Yang, W.-Q. Wang, and S. Zhang, ``Two-dimensional direction estimation of multiple signals using two parallel sparse linear arrays," \emph{
Signal Process.}, vol. 143, pp. 112--121, Feb. 2018.
\bibitem{Lshaped1}
Z. Zheng and S. Mu, ``2-D direction finding with pair-matching operation for L-shaped nested array," \emph{IEEE Commun. Lett.}, vol. 25, no. 3, pp. 975--979, Mar. 2021.
\bibitem{Lshaped2}
Y. Yang, X. Mao, Y. Hou, and G. Jiang, ``2-D DOA estimation via correlation matrix reconstruction for nested L-shaped array," \emph{Digit. Signal
Process.}, vol. 98, pp. 102623--102633, Mar. 2020.
\bibitem{nonparallel1}
A. M. Elbir, ``V-shaped sparse arrays for 2-D DOA estimation," \emph{
Circuits, Syst., Signal Process.}, vol. 38, no. 6, pp. 2792--2809, Jun. 2019.
\bibitem{nonparallel2}
X. Wu and W.-P. Zhu, ``Single far-field or near-field source localization
with sparse or uniform cross array," \emph{IEEE Trans. Veh. Technol.}, vol. 69, no. 8, pp. 9135--9139, Aug. 2020.
\bibitem{planar}
Q. Wu, F. Sun, P. Lan, G. Ding, and X. Zhang, ``Two-dimensional direction-of-arrival estimation
for co-prime planar arrays: A partial spectral search approach," \emph{IEEE Sensors J.}, vol. 16, no. 14, pp. 5660--5670, Jul. 2016.
\bibitem{NA2D1}
P. Pal and P. P. Vaidyanathan, ``Nested arrays in two dimensions, Part I: Geometrical considerations," \emph{IEEE Trans. Signal Process.}, vol. 60,
no. 9, pp. 4694--4705, Sep. 2012.
\bibitem{NA2D2}
P. Pal and P. P. Vaidyanathan, ``Nested arrays in two dimensions, Part II: Application in two dimensional array processing," \emph{IEEE Trans.
Signal Process.}, vol. 60, no. 9, pp. 4706--4718, Sep. 2012.
\bibitem{other1}
L. Sun, M. Yang, and B. Chen, ``Thermos array: Two-dimensional sparse
array with reduced mutual coupling," \emph{Int. J. Antennas Propag.}, vol. 2018, pp. 1--8, Jan. 2018.
\bibitem{other2}
R. Rajamki and V. Koivunen, ``Sparse active rectangular array with few
closely spaced elements," \emph{IEEE Signal Process. Lett.}, vol. 25, no. 12, pp. 1820--1824, Dec. 2018.
\bibitem{minhongqi}
H. Min, C. Feng, R. Li, and Y. Zeng, ``Integrated sensing and communication with nested array: Beam pattern and performance analysis," \emph{in Proc. Int. Conf. Wirel. Commun. Signal Process. (WCSP)}, 2024, pp. 764--769.
\bibitem{R.Long}
R.~Long, Y.-C.~Liang, H.~Guo, G.~Yang, and R.~Zhang, ``Symbiotic radio: A new communication paradigm for passive Internet-of-Things," \emph{IEEE Internet Things J.}, vol.~7, pp.~1350--1363, Nov.~2020.
%\bibitem{2DNA1}
%P. Pal and P. P. Vaidyanathan, ``Nested arrays in two dimensions, Part
%I: Geometrical considerations," \emph{IEEE Trans. Signal Process.}, vol. 60,
%no. 9, pp. 4694--4705, Sep. 2012.
%\bibitem{2DNA2}
%P. Pal and P. P. Vaidyanathan, ``Nested arrays in two dimensions, Part
%II: Application in two dimensional array processing," \emph{IEEE Trans.
%Signal Process.}, vol. 60, no. 9, pp. 4706--4718, Sep. 2012.
\bibitem{musicDEDI}
R.~Schmidt, ``Multiple emitter location and signal parameter estimation," \emph{IEEE Trans. Antennas Propag.}, vol.~AP-34, no.~3, pp.~276--280,
Mar.~1986.
\bibitem{wcnc}
J. Xu, H. Wang, Y. Zeng and X. Xu, ``Little pilot is needed for channel estimation
with integrated super-resolution sensing and communication," \emph{in Proc. IEEE Wireless Commun. and Networking Conf.(WCNC)}, 2024, pp. 1--6.
\end{thebibliography}
\end{document}